\documentclass[aps,prx,12pt,letterpaper,amsmath,amssymb]{revtex4-1}

\usepackage[utf8]{inputenc}
\usepackage{graphicx}
\usepackage{subfigure}
\usepackage[usenames,dvipsnames]{xcolor}
\usepackage{epstopdf}
\usepackage{amsmath,amsthm,amssymb}
\usepackage{latexsym}
\usepackage{bm}
\usepackage{dcolumn}
\usepackage{braket}
\usepackage[normalem]{ulem}
\usepackage{times}
\usepackage{hyperref}
\usepackage{enumitem}

\newcommand{\ignore}[1]{}
\newcommand{\red}[1]{\textcolor{black}{#1}}

\newcommand{\boxedpph}{\boxed{++}}
\newcommand{\boxedmmh}{\boxed{--}}
\newcommand{\boxedpmh}{\boxed{+-}}
\newcommand{\boxedmph}{\boxed{-+}}
\newcommand{\boxedpp}{\boxed{{}_+^+}}
\newcommand{\boxedpm}{\boxed{{}_+^-}}
\newcommand{\boxedmp}{\boxed{{}_-^+}}
\newcommand{\boxedmm}{\boxed{{}_-^-}}

\begin{document}
\bibliographystyle{naturemag}

\title{\red{A paradigm of} spontaneous marginal phase transition at\\ finite temperature in one-dimensional ladder Ising models}
\author{Weiguo Yin}
\email{wyin@bnl.gov}
\affiliation{Condensed Matter Physics and Materials Science Division,
Brookhaven National Laboratory, Upton, New York 11973, USA}

\begin{abstract}
The Ising model 
describes collective behaviors such as phase transitions and critical phenomena in various physical, biological, economical, and social systems. It is well-known that spontaneous phase transition at finite temperature does not exist in the Ising model with short-range interactions in one dimension. \red{Yet, little is known about whether this forbidden phase transition can be approached arbitrarily closely---at fixed finite temperature. 
To describe such asymptoticity, here I introduce the notion of marginal phase transition~(MPT) and
use symmetry analysis of the transfer matrix to reveal the existence of spontaneous MPT at fixed finite temperature $T_0$ in one class of one-dimensional Ising models on decorated two-leg ladders, in which $T_0$ is determined solely by on-rung interactions and decorations, while the crossover width $2\delta T$ is independently, exponentially reduced ($\delta T=0$ means a genuine phase transition) by on-leg interactions and decorations. These findings establish a simple ideal paradigm for realizing an infinite number of one-dimensional Ising systems with spontaneous MPT, which would be characterized in routine lab measurements as a genuine first-order phase transition with large latent heat thanks to the ultra-narrow $\delta T$ (say less than one nano-kelvin), paving a way to push the limit in our understanding of phase transitions and the dynamical actions of frustration arbitrarily close to the forbidden regime.}


\end{abstract}

{
}
\date{\today}

\maketitle

\section{Introduction}

The textbook Ising model~\cite{Mattis_book_08_SMMS,Mattis_book_1985,Baxter_book_Ising,003_Huang_08_book} consists of individuals that have one of two values ($+1$ or $-1$, e.g.,  magnetic moments of atomic spins pointing to the up or down direction, open or close in neural networks, yes or no in voting, etc.): 
\begin{eqnarray}
H=-\sum_{ij}^{}{J_{ij}\sigma_{i}\sigma_{j}} - h\sum_{i}^{}\sigma_{i},
\end{eqnarray}
where $\sigma_{i}=\pm 1$ is the $i$th individual's value and $h$ depicts the bias (e.g., the external magnetic field that favors $\sigma_{i}=+1$ for $h>0$). The individuals interact according to the simple rule that neighbors with like values are rewarded more than those with unlike values (i.e., the ferromagnetic interaction $J_{ij}>0$).
Therefore, the society tends to form the order in which all the members have the same value. This tendency is however disturbed by heat, which favors the free choice of the values or disorder. The model was originally proposed for a central question in condensed matter physics, namely whether a \emph{spontaneous} phase transition (i.e., in the absence of the bias, $h=0$) between the high-temperature disordered state and the low-temperature ordered state exists at finite temperature, and in his 1924 doctoral thesis\red{---one century ago---}Ernst Ising solved the model exactly for spins localized on a one-dimensional (1D) single chain but found no phase transition~\cite{Ising1925}. The 2D square-lattice Ising model, which is much harder to be solved exactly, is generally referred to as one of the simplest statistical models to show a spontaneous phase transition at finite temperature~\cite{Onsager_Ising_2D}. Moving from one to two dimension was often described with $n$-leg spin ladders (i.e., $n$ coupled parallel chains, which are still 1D; the $n\to\infty$ limit is 2D)~\cite{Dagotto_science_ladder}, and again no phase transition was found for finite $n$~\cite{Mejdani_Ising_ladder}.
In fact, the Perron–Frobenius theorem implied the nonexistence of any phase transition in 1D Ising models with short-range interactions~\cite{Cuesta_1D_PT}. 
\red{The rigorously proved nonexistence of spontaneous phase transition at finite temperature in the 1D Ising model is deeply rooted in symmetry~\cite{Batista_PRB_05_dimensional-reductions} and becomes one of the most quoted limits of human knowledge.}

\red{Here, we ask the question of how close the 1D Ising model can get to a phase transition. The ideal answer would be such an ultra-narrow phase crossover that can approach a genuine phase transition arbitrarily closely in a definite way---at a fixed finite temperature $T_0$ \emph{and} with the crossover width $2\delta T$ getting narrower and narrower, best exponentially ($\delta T=0$ means phase transition). We coin the term ``marginal phase transition'' (MPT) for phase crossovers with the above asymptoticity. Thus, MPT means a pathway in which broad phase crossovers can be readily tuned to be ultra-narrow at the same $T_0$. Practically speaking, when $\delta T$ becomes so narrow (say less than one nano-kelvin), the MPT would be characterized as a genuine phase transition in lab measurements. The quest for MPT will push the limit in our understanding of phase transitions as close as possible to the forbidden regime. 
}

\red{A recent breakthrough relevant to this quest was the discovery of the so-called ``pseudo-transition'' in a few decorated Ising chains in the presence of the magnetic field~\cite{005_Galisova_PRE_15_double-tetrahedral-chain,
007_Torrico_PRA_16_Ising-XYZ-diamond-chain,
009_review_Souza_SSC_18_Ising-XYZ-diamond-chain_double-tetrahedral-chain-spin-electron,
010_Carvalho_JMMM_18_Ising-XYZ-diamond-chain_quantum-entanglement,
011_Rojas_BJP_20_Ising-Heisenberg-tetrahedral_diamond,
013_Rojas_PRE_19_previous_4_models,
014_Rojas_JPC_20_Ising–Heisenberg_spin-1-double-tetrahedral-chain,
015_Strecka_APPA_20_Ising-diamond-chain,
015-7_Canova_CzechoslovakJP_04_Ising–Heisenberg_diamond_chain,
015-8_Canova_JPC_06_Ising–Heisenberg-spin-S-diamond-chain,
017_Krokhmalskii_PA_21_3-previous-chains_effective_model,
016_Strecka_book_chapter}. For $h\ne 0$, those 1D Ising chains exhibit the entropy jump and a gigantic peak in specific heat in a narrow temperature region, resembling a first-order phase transition. The construction of the ``pseudo-transition'' utilized the concept of frustration as follows:} The Ising model can also accommodate the opposite rule, namely neighbors with like values are rewarded \emph{less} than those with unlike values (i.e., antiferromagnetic interaction  $J_{ij}<0$). Accordingly, the ordered state has alternating values if such an arrangement can be made, such as $+1$/$-1$/$+1$/$-1$/$+1$/$-1\cdots$ on a 1D chain lattice, the checkerboard pattern on the 2D square lattice, etc. (for bipartite lattices, the antiferromagnetic Ising model is equivalent to the ferromagnetic one by the transformation of $\sigma_i \to -\sigma_i$ in one of the two sublattices). 
However, there are numerous situations where the between-neighbor interactions cannot be simultaneously satisfied. The standard example is a spin triangle (Fig.~\ref{Fig:model}a) where two spins have opposite values, then it is impossible for the third spin to have the opposite value to both of them, leading to the state degeneracy of 2. This phenomenon is called \emph{geometrical frustration}~\cite{Balents_nature_frustration,Miyashita_10_review_frustration}. \red{Meanwhile, if the non-degenerate state, where the first two spins couple more strongly to the third spin and satisfactorily have the same value (because of having the opposite value to the third spin), is tuned to be the ground state with slightly lower energy than the aforementioned degenerated state, then a thermal (entropy) driven crossover between them would occur at finite temperature~\cite{Miyashita_10_review_frustration}. This physics of phase crossover is rather generic; what is challenging is how to make the crossover ultra-narrow and asymptotic with fixed $T_0$. It was significant that the ``pseudo-transition'' has showcased the existence of ultra-narrow phase crossover in the 1D Ising model for $h \ne 0$. For $h=0$, however, the pseudo-transition in those decorated Ising chains is prohibited by spin up-down symmetry~\cite{005_Galisova_PRE_15_double-tetrahedral-chain,
007_Torrico_PRA_16_Ising-XYZ-diamond-chain,
009_review_Souza_SSC_18_Ising-XYZ-diamond-chain_double-tetrahedral-chain-spin-electron,
010_Carvalho_JMMM_18_Ising-XYZ-diamond-chain_quantum-entanglement,
011_Rojas_BJP_20_Ising-Heisenberg-tetrahedral_diamond,
013_Rojas_PRE_19_previous_4_models,
014_Rojas_JPC_20_Ising–Heisenberg_spin-1-double-tetrahedral-chain,
015_Strecka_APPA_20_Ising-diamond-chain,
015-7_Canova_CzechoslovakJP_04_Ising–Heisenberg_diamond_chain,
015-8_Canova_JPC_06_Ising–Heisenberg-spin-S-diamond-chain,
017_Krokhmalskii_PA_21_3-previous-chains_effective_model,
016_Strecka_book_chapter}, suggesting that spontaneous ultra-narrow phase crossover be much harder to realize.} 
\red{We noticed that three cases of zero-field ``pseudo-transition'' were reported to occur in 2-leg~\cite{016_Strecka_book_chapter,008_zero-field_Rojas_SSC_16_Ising-Heisenberg_ladder} and 3-leg~\cite{006_zero-field_Strecka_JMMM_16_Ising-Heisenberg-3-leg-tube} ladders; however, the insight into how these few 1D Ising models can escape from the symmetry constraint for the single chain is lacking---as shown below, such symmetry analysis is vital to the generation of an infinite number of zero-field ultra-narrow phase crossover.
Above all, the asymptoticity required by MPT has not been addressed in the context of ``pseudo-transition,'' where even the crossover width $2\delta T$ was not clearly defined and rigorously expressed in terms of the model parameters~\cite{005_Galisova_PRE_15_double-tetrahedral-chain,
007_Torrico_PRA_16_Ising-XYZ-diamond-chain,
009_review_Souza_SSC_18_Ising-XYZ-diamond-chain_double-tetrahedral-chain-spin-electron,
010_Carvalho_JMMM_18_Ising-XYZ-diamond-chain_quantum-entanglement,
011_Rojas_BJP_20_Ising-Heisenberg-tetrahedral_diamond,
013_Rojas_PRE_19_previous_4_models,
014_Rojas_JPC_20_Ising–Heisenberg_spin-1-double-tetrahedral-chain,
015_Strecka_APPA_20_Ising-diamond-chain,
015-7_Canova_CzechoslovakJP_04_Ising–Heisenberg_diamond_chain,
015-8_Canova_JPC_06_Ising–Heisenberg-spin-S-diamond-chain,
017_Krokhmalskii_PA_21_3-previous-chains_effective_model,
016_Strecka_book_chapter,008_zero-field_Rojas_SSC_16_Ising-Heisenberg_ladder,006_zero-field_Strecka_JMMM_16_Ising-Heisenberg-3-leg-tube}. 
}



The purpose of this paper is to present the  finding of spontaneous MPT (SMPT) in a family of decorated Ising 2-leg ladders with strong frustration\red{~\cite{Yin_MPT}, in which $T_0$ is determined by on-rung interactions, while $\delta T$ is independently, exponentially reduced by on-leg interactions for fixed $T_0$.} This establishes a simple ideal paradigm for implementing an infinite number of 1D Ising systems with SMPT~\cite{Yin_icecreamcone}. We further found that the SMPT can be expressed \red{accurately and conveniently} by a nonclassical order parameter providing a \red{microscopic} description of the abrupt switching between two unconventional long-range orders where the local two-parent-spin correlations are ferromagnetic and antiferromagnetic, respectively. 
\red{Moreover, we show that the on-leg decoration (which controls $\delta T$) can be done independently of the on-rung decoration (which controls $T_0$), revealing an amazing advantage of this paradigm. Specifically, $\delta T$ can be exponentially reduced by increasing the number of on-leg-decorated spins, making it clear that spontaneous phase transition at finite temperature does not exist in the 1D Ising model with \emph{a finite number of} short-range interactions.}


\red{
The rest of the paper is organized as follows: Section~\ref{strategy} heuristically describes the line of thinking and the use of a dimensionality increase and reduction method that utilized symmetry analysis to give rise to the SMPT. For the sake of clarity, Section~\ref{trimer} details the realization of SMPT in a minimal 1D Ising model by decorating the rungs and the exact solutions about its thermodynamic properties, correlation functions, and nonclassical order parameters. Section~\ref{asymptotic} showcases how to decorate the legs to approach a genuine phase transition at finite temperature arbitrarily closely. Section~\ref{discussion} addresses some immediate implications of the present work on further fundamental and technological research and development.
}

\red{
\section{The Method}
\label{strategy}
}

\subsection{SMPT from mimicking the first-order phase transition}
The mathematical signature of phase transitions is the non-analyticity of the system's thermodynamic free energy $f(T)$ where $T$ denotes temperature. A $k$th-order phase transition means that the $k$th derivative of $f(T)$ starts to be discontinuous at the transition. The exactly solved 1D Ising models on both the single chain~\cite{Mattis_book_08_SMMS,Mattis_book_1985,Baxter_book_Ising,003_Huang_08_book} and ordinary $n$-leg ladders with $n=2,3,4$~\cite{Mejdani_Ising_ladder} offer examples of analytic free energies. 

\begin{figure}[t]
    \begin{center}
        \subfigure[][]{
\includegraphics[width=0.48\columnwidth,clip=true,angle=0]{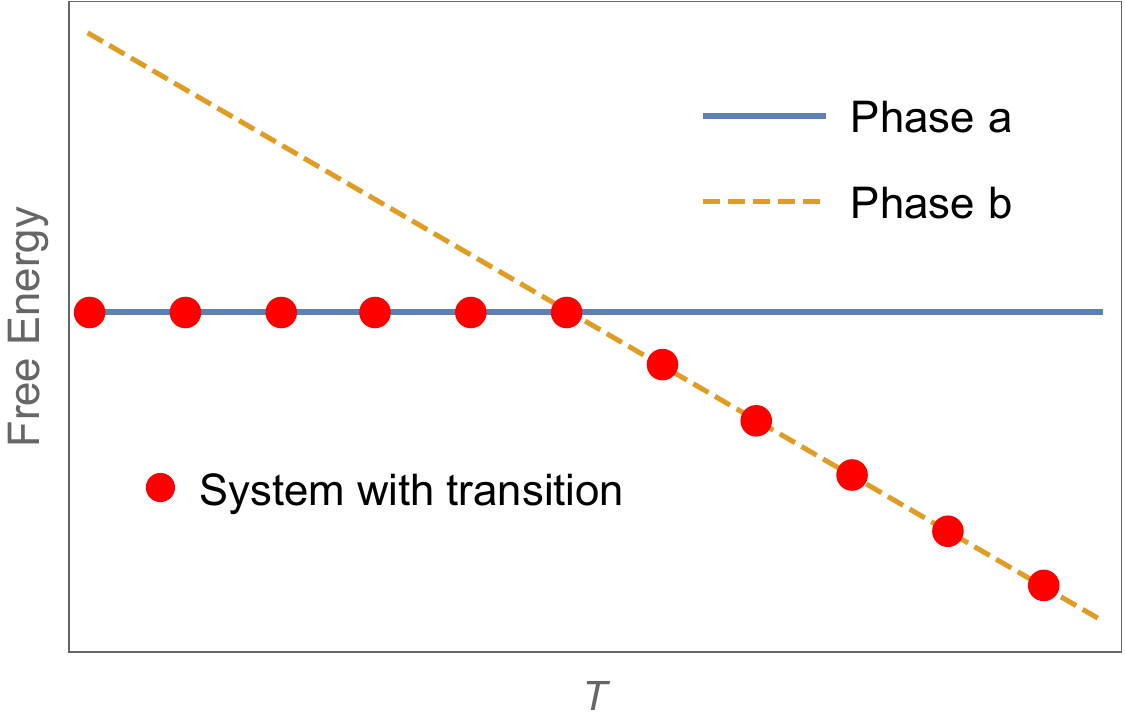}
        }
        \subfigure[][]{
\includegraphics[width=0.48\columnwidth,clip=true,angle=0]{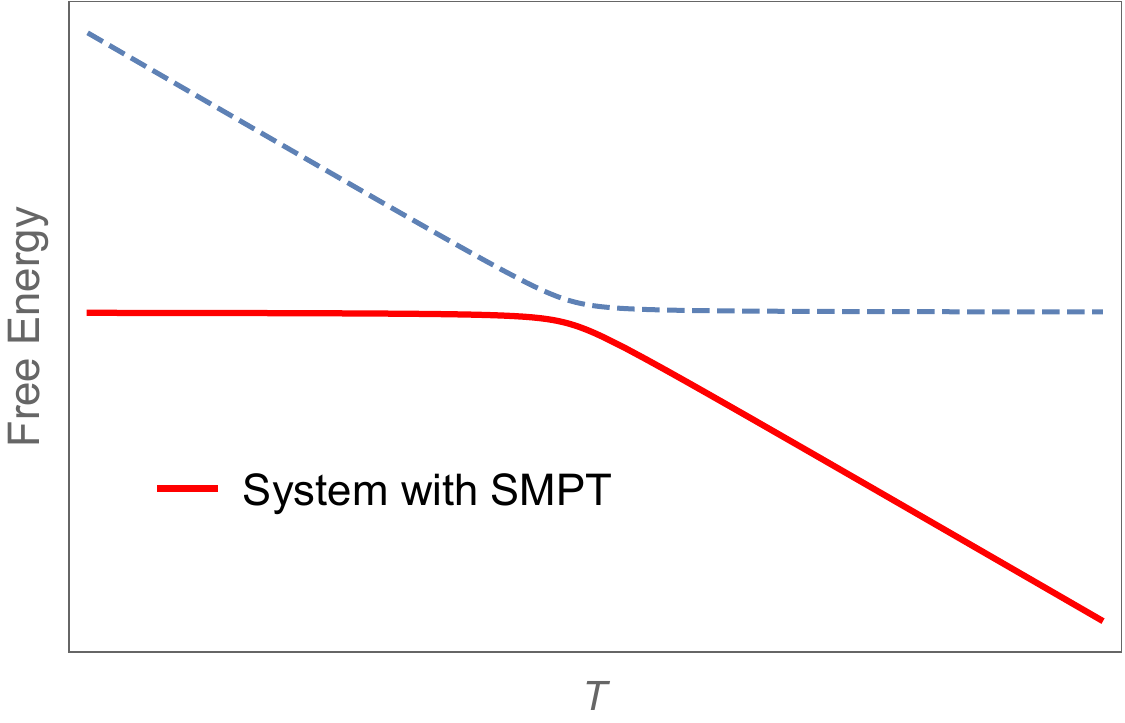}
        }
    \end{center}
\caption{\textbf{SMPT mimics a genuine phase transition.} (a) Schematic of a first-order phase transition due to the free-energy crossing of the two phases a and b (lines). The free energy of the whole system is represented by the line (not shown) connecting the red dots and is non-analytic at the level crossing. (b) Schematic of SMPT, whose free energy looks the same as (a) except when getting too close to the ultra-narrow phase crossover. For shorthand notation, $T$ means $k_\mathrm{B}T$ in all the figures labels from now on.}
\label{Fig:thinking}
\end{figure}

The simplest description of the phase transition phenomenon is the level crossing in the first-order phase transition, such as melting of ice or the boiling of water. As illustrated in Fig.~\ref{Fig:thinking}a, 
suppose the free-energy functions of two phases $-g[a(T)]$ and $-g[b(T)]$ cross at the critical temperature $T_0$. Then, the free energy of the system taking the lower value of $-g[a(T)]$ and $-g[b(T)]$ is given by
\begin{eqnarray}
f_\mathrm{PT}(T)=-g\left[\frac{a+b}{2}+\frac{|a-b|}{2}\right],
\label{fPT}
\end{eqnarray}
which is nonanalytic at $T_0$ with its first derivative with respect to $T$ being discontinuous. 

We hypothesize that SMPT can be created by using an analytic function to mimic Eq.~(\ref{fPT}) and consider here~\cite{009_review_Souza_SSC_18_Ising-XYZ-diamond-chain_double-tetrahedral-chain-spin-electron} 
\begin{eqnarray}
f_\mathrm{SMPT}(T)=-g\left[\frac{a+b}{2}+\sqrt{\left(\frac{a-b}{2}\right)^2+c^2}\right],
\label{fSMPT}
\end{eqnarray}
\textbf{which satisfies the following two conditions:} 
(i) $a(T)$ and $b(T)$ cross at $T_0$, and (ii) $c(T) \ll |a-b|$ except for an ultra-narrow temperature region around $T_0$. The width of the crossover $2\delta T$ can be estimated by solving $\frac{|a-b|}{2}=c$ at $T=T_0 \pm \delta T$. This definition of $\delta T$ is consistent with the crossover width later defined by measuring the order parameter [see Eq.~(\ref{dT2a})].  
Eq.~(\ref{fSMPT}) looks the same as Eq.~(\ref{fPT}) except when zoomed into the ultra-narrow region around $T_0$ (see Fig.~\ref{Fig:thinking}b). 

The form of Eq.~(\ref{fSMPT}) can result from the 1D Ising model on a single chain. In the thermodynamic limit, the free energy per unit cell $f(T)=-\lim^{}_{N\to\infty}\frac{1}{N} k_\mathrm{B}T\ln Z$, where $N$ is the number of unit cells, $Z=\mathrm{Tr}\left( e^{-\beta H} \right)$ is the partition function, 
and  $\beta=1/(k_\mathrm{B}T)$ with 
$k_\mathrm{B}$ being the Boltzmann constant.  
$Z$ of a 1D Ising model can be obtained exactly by using the transfer matrix method~\cite{Mattis_book_08_SMMS,Mattis_book_1985,Baxter_book_Ising,003_Huang_08_book,Mejdani_Ising_ladder,005_Galisova_PRE_15_double-tetrahedral-chain,
007_Torrico_PRA_16_Ising-XYZ-diamond-chain,
009_review_Souza_SSC_18_Ising-XYZ-diamond-chain_double-tetrahedral-chain-spin-electron,
010_Carvalho_JMMM_18_Ising-XYZ-diamond-chain_quantum-entanglement,
011_Rojas_BJP_20_Ising-Heisenberg-tetrahedral_diamond,
013_Rojas_PRE_19_previous_4_models,
014_Rojas_JPC_20_Ising–Heisenberg_spin-1-double-tetrahedral-chain,
015_Strecka_APPA_20_Ising-diamond-chain,
015-7_Canova_CzechoslovakJP_04_Ising–Heisenberg_diamond_chain,
015-8_Canova_JPC_06_Ising–Heisenberg-spin-S-diamond-chain,
017_Krokhmalskii_PA_21_3-previous-chains_effective_model,
016_Strecka_book_chapter,008_zero-field_Rojas_SSC_16_Ising-Heisenberg_ladder,006_zero-field_Strecka_JMMM_16_Ising-Heisenberg-3-leg-tube,Yin_MPT,Yin_icecreamcone} 
and is given by 
\begin{equation}
Z=\mathrm{Tr}\left(\Lambda^N\right)=\sum_k{\lambda_k^N} \;\;\to\;\; \lambda^N \;\mathrm{for}\; N\to \infty,
\end{equation} 
where $\Lambda$ is the transfer matrix, $\lambda_k$ the $k$th eigenvalue of $\Lambda$, and $\lambda$ the largest eigenvalue. Thus, $f(T)=-k_\mathrm{B}T\ln\lambda$. For the single-chain Ising model, the transfer matrix is of the following form:
\begin{eqnarray}
\Lambda_\mathrm{single-chain}=\left(
\begin{array}{cc}
  a & c \\
  c & b \\
\end{array}
\right),
\label{single}
\end{eqnarray}
with $\lambda=\frac{1}{2}\big[a+b+\sqrt{(a-b)^2+4c^2}\big]$. Compared with Eq.~(\ref{fSMPT}), $f(T)$ is of the same form when the function $g(x)$ is defined as $g(x)=k_\mathrm{B}T\ln(x)$.

\subsection{The absence of SMPT in single-chain Ising models}
However, in the absence of the bias ($h=0$), the Ising model is invariant with respect to flipping all $\sigma_i$. By this spin up-down symmetry, $a=b$ and $\lambda=a+c$ in Eq.~(\ref{single}) for single-chain Ising models, including decorated Ising chains. 
This means the absence of SMPT, which requires the crossing of $a$ and $b$ as $T$ changes. 

\ignore{
Specifically, for the simplest single-chain model depicted in Fig.~2a, $a=b=e^{\beta J}$, $c=e^{-\beta J}$, and $\lambda=2\cosh{\beta J}$. There is no phase crossover at all. 
Whereas, for decorated single-chain models, frustration could give rise to a broad phase crossover due to the crossing of $a$ and $c$. For example, for the decorated model depicted in Fig.~2d, $a=b=2e^{\beta J}\cosh(2\beta J')$ and $c=2e^{-\beta J}$. The broad crossover takes place when $a=c$ at $T_0\approx \frac{2}{k_\mathrm{B}\ln 2} (|J'|+J)$ for $|J'|>-J>0$.

\subsection{Pseudo phase transition in the presence of the bias}
}

The nonzero bias $h$ breaks the above invariance and lifts the degeneracy of $a$ and $b$. For the simplest single-chain model, 
$a=e^{\beta (J+h)}$, $b=e^{\beta (J-h)}$, $c=e^{-\beta J}$. Thus, $a$ and $b$ do not cross at all. 
How to make $a$ and $b$ cross in decorated single-chain Ising models in the presence of the bias was the subject of the so-called ``pseudo transition''~\cite{005_Galisova_PRE_15_double-tetrahedral-chain,
007_Torrico_PRA_16_Ising-XYZ-diamond-chain,
009_review_Souza_SSC_18_Ising-XYZ-diamond-chain_double-tetrahedral-chain-spin-electron,
010_Carvalho_JMMM_18_Ising-XYZ-diamond-chain_quantum-entanglement,
011_Rojas_BJP_20_Ising-Heisenberg-tetrahedral_diamond,
013_Rojas_PRE_19_previous_4_models,
014_Rojas_JPC_20_Ising–Heisenberg_spin-1-double-tetrahedral-chain,
015_Strecka_APPA_20_Ising-diamond-chain,
015-7_Canova_CzechoslovakJP_04_Ising–Heisenberg_diamond_chain,
015-8_Canova_JPC_06_Ising–Heisenberg-spin-S-diamond-chain,
017_Krokhmalskii_PA_21_3-previous-chains_effective_model,
016_Strecka_book_chapter}. It was understood that certain decorations could renormalize the bias to be effectively $T$-dependent and the resulting $h_\mathrm{eff}(T)$ changes its sign at the pseudo-critical temperature $T_0$~\cite{017_Krokhmalskii_PA_21_3-previous-chains_effective_model}. 
Nevertheless, $h_\mathrm{eff}(T)$ vanishes for $h=0$, in agreement with the aforementioned spin up-down symmetry. Therefore, no spontaneous pseudo-transition would occur in those systems.

The extensively studied 1D Ising chain with the first- and second-nearest-neighbor interactions, the $J_1$-$J_2$ model~\cite{Fleszar_Baskaran_JPC_85_J1-J2,Stephenson_CanJP_70_J1-J2-Ising-chain,Dobson_JMathP_69_Many‐Neighbored-Ising-Chain,Dhar_PRE_00_J1-J2-Ising-chain}, 
showcases the emerging of an effective magnetic field $h_\mathrm{eff}$ for $h=0$ by the transformation $\sigma_i\sigma_{i+1} = \Tilde{\sigma}_i$. This was inspiring. However, $h_\mathrm{eff}=J_1$, the nearest-neighbor interaction, is not $T$-dependent, and the transformed model is an ordinary single-chain Ising model in this $T$-independent effective field~\cite{Fleszar_Baskaran_JPC_85_J1-J2,Stephenson_CanJP_70_J1-J2-Ising-chain,Dobson_JMathP_69_Many‐Neighbored-Ising-Chain}. Therefore, the resultant effective $a$ and $b$ do not cross at all.

\subsection{Dimensionality increase and reduction for SMPT}

To find SMPT, we resort to the strategy of dimensionality increase and reduction: First, we increase the transfer matrix from $2\times 2$ to $4\times 4$. Then, we identify the $4\times 4$ transfer matrix that has such high symmetry that it can be block diagonalized and reduced back to a $2\times 2$ matrix~\cite{Mattis_book_08_SMMS} now with the new effective $a$ and $b$ crossing. 

We consider the following $4\times 4$ transfer matrix
\begin{eqnarray}
\Lambda=\left(
\begin{array}{cccc}
 a & z & z & u \\
 z & b & v & z \\
 z & v & b & z \\
 u & z & z & a \\
\end{array}
\right)
\label{TM4simple}
\end{eqnarray}
It can be block diagonalized by the parity-symmetry operations $U$: 
\begin{eqnarray}
U=\frac{1}{\sqrt{2}}\left(
\begin{array}{cccc}
 1 & 0 & 0 & 1 \\
 0 & 1 & 1 & 0 \\
 0 & -1 & 1 & 0 \\
 -1 & 0 & 0 & 1 \\
\end{array}
\right)
\label{U}
\end{eqnarray}
and the result is
\begin{eqnarray}
U^T \Lambda U=\left(
\begin{array}{cccc}
 a-u & 0 & 0 & 0 \\
 0 & b-v & 0 & 0 \\
 0 & 0 & b+v & 2z \\
 0 & 0 & 2z & a+u \\
\end{array}
\right)
\label{transformed}
\end{eqnarray}
\ignore{where $a+u=2\cosh(2x+2x')e^w\boxedpp^2$, $b+v=2\cosh(2x-2x')e^{-w}\boxedpm^2$,
$a-u=2\sinh(2x+2x')e^w\boxedpp^2$,
$b-v=2\sinh(2x-2x')e^{-w}\boxedpm^2$, and
$z=\boxedpp\boxedpm$.
The eigensystem problem is reduced to a $2\times2$ matrix problem for the even-parity states [the bottom right part of Eq.~(\ref{transformed})], i.e.,
\begin{eqnarray}
(U^T \Lambda U)_\mathrm{even-parity}=\left(
\begin{array}{cc}
 b+v & 2z \\
  2z & a+u \\
\end{array}
\right)
\label{2x2}
\end{eqnarray}}
Thus, the eigenvalues of the $4\times 4$ transfer matrix are $a-u$, $b-v$, and
\begin{equation}
\lambda_\pm=\frac{a+u+b+v}{2} \pm  \sqrt{\left(\frac{a+u-b-v}{2}\right)^2+4z^2}.
\label{general}
\end{equation}
Then, the task is transformed to how to realize the crossing of $a+u$ and $b+v$.

\red{The $4\times 4$ transfer matrix can be obtained for the 1D Ising model on a 2-leg ladder. The ordinary 2-leg Ising ladder~\cite{Mejdani_Ising_ladder}, the spin-1/2 Ising tetrahedral chain~\cite{016_Strecka_book_chapter}, and the spin-1/2 Ising-Heisenberg ladder with alternating Ising and Heisenberg inter–leg couplings~\cite{008_zero-field_Rojas_SSC_16_Ising-Heisenberg_ladder} are known special cases satisfying the symmetry of Eq.~(\ref{TM4simple}). Among them, the ordinary 2-leg ladder does not host the SMPT, and the latter two cases with zero-field pseudo-transition have not been analyzed with the above symmetry-based block diagonalization technique, which is vital to systematically looking for more cases of spontaneous ultra-narrow phases crossover. Here, we take the ordinary 2-leg ladder as the parent system and investigate how to properly decorate it with child spins.}

\section{A minimal model}
\label{trimer}

\begin{figure}[t]
\vspace{-1cm}
\includegraphics[width=0.7\columnwidth,clip=true,angle=0]{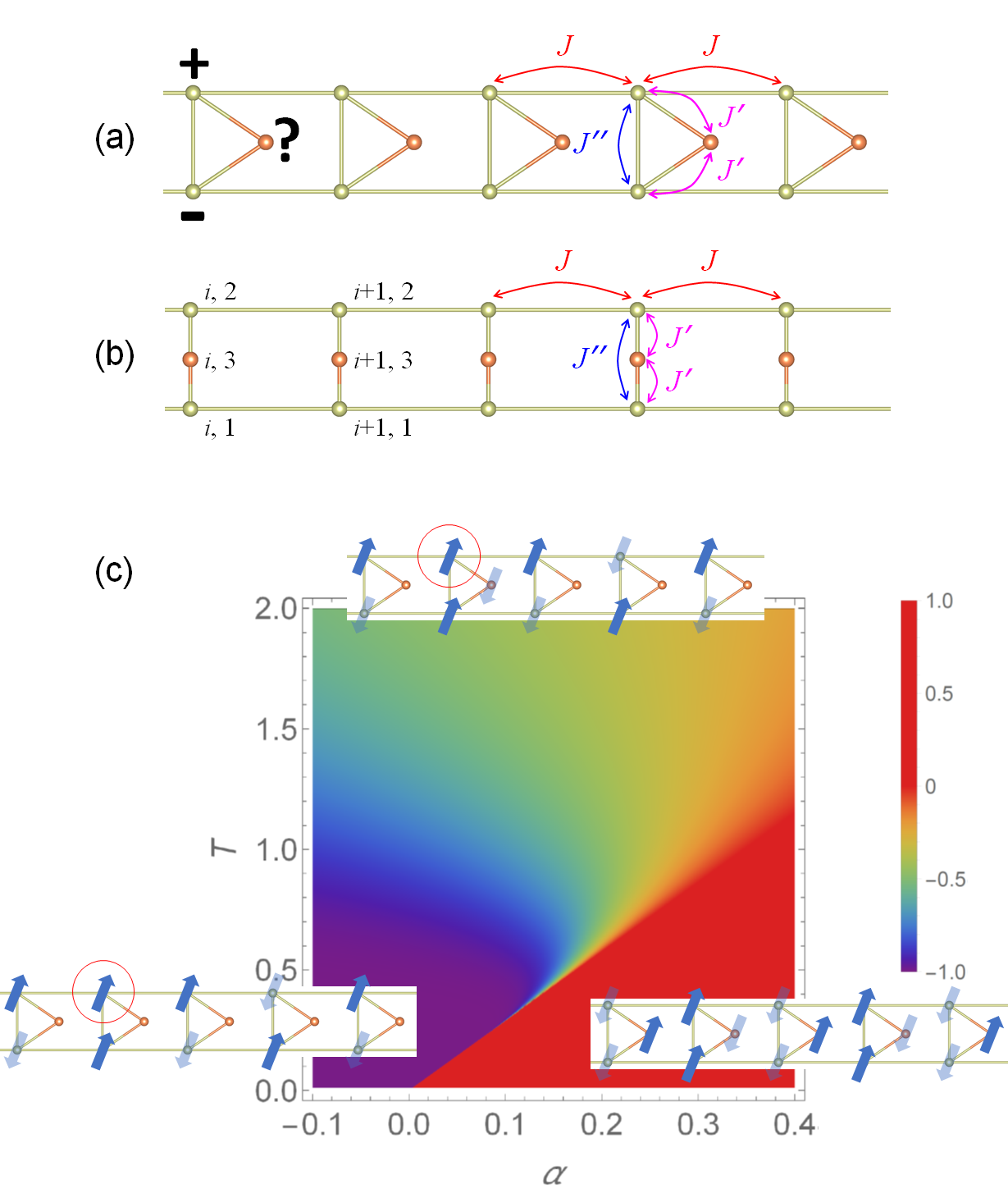}
\caption{\textbf{The model and the phase diagram.} The 2-leg ladder with (a) triangle rungs or (b) trimer rungs, where the balls stand for the spins and the bonds for the interactions $J$, $J'$, and $J''$. The two ladders are equivalent. Spin frustration is illustrated in the leftmost triangle rung of (a): when the two end spins have opposite values, the middle one has no preference despite strong interactions. (c) The color map of the order parameter $C_{12}(0)$ as a function of temperature $T$ and the frustration parameter $\alpha=(|J'|+J'')/|J|$. $J''=-3$ and $|J|=1$ is the energy unit. Red stands for the nearly $+1$ region governed by $J'$ (where spin1 and spin 2 on the same rung have like values), purple stands for the nearly $-1$ region governed by $J''$ (where spin 1 and spin 2 on the same rung have unlike values and spin 3 is decoupled). The SMPT between them takes place at $T_0=2(|J'|+J'')/(k_\mathrm{B}\ln2)$ with the ultra-narrow crossover width $2\delta T = 2^{-1/\alpha}\,T_0\,8/\ln{2}$  for strong frustration $0<\alpha < 0.15$. The bluish-to-greenish region is the exotic high-$T$ phase where frustration effectively decouples the two legs of the ladder. \textbf{The microscopic mechanism of the decoupling} is illustrated with two kinds of spin flip, which are marked by the red circles: The spin flip that does not involve the child spin (bottom left) costs the energy of $4|J|+2|J''|$. The spin flip in the high-$T$ phase (top), for which the child spin responds cooperatively, costs the energy of $4|J|+2|J''|-2|J'| \approx 4|J|$, similar to the case of two decoupled chains. In the illustrations, $J<0$ and $J'<0$ are used without loss of generality.}
\label{Fig:model}
\end{figure}

For clarity, we focus the presentation on the trimer-rung ladder that has only one decorated spin per rung in this section. The spin trimer (Fig.~\ref{Fig:model}b) or equivalently triangular (Fig.~\ref{Fig:model}a; hence the source of frustration is more obvious) represents the simplest form of frustration. The model is defined as
\begin{eqnarray}
H&=&H^{}_\mathrm{parents}+\sum_{i} H^{(i)}_\mathrm{children}, \label{minimal} \\
H^{}_\mathrm{parents}&=&-\sum_{i=1}^{N}{\left[J(\sigma_{i,1}\sigma_{i+1,1}+\sigma_{i,2}\sigma_{i+1,2})
+J''\sigma_{i,1}\sigma_{i,2}\right]}, \nonumber \\ 
H^{(i)}_\mathrm{children}&=&-J'(\sigma_{i,1}\sigma_{i,3}+\sigma_{i,2}\sigma_{i,3}),
\nonumber 
\end{eqnarray}
where $\sigma_{i,m}=\pm 1$ denotes the spins on the $m$th site of the $i$th rung of the ladder and $\sigma_{N+1,m}\equiv\sigma_{1,m}$ (i.e., the periodic boundary condition). $N$ is the total number of the rungs and we are interested in the thermodynamical limit $N\to\infty$. Each rung has three sites with $m=1,2$ on the two legs \red{(referred to as the \emph{parent} sites)} and $m=3$ decorated at the middle \red{(referred to as the \emph{child} site)} of the rung; \red{so a rung is also called a \emph{household} for ease of memory.} $J$ is the nearest-neighbor interaction along the legs, $J''$ the interaction that directly couples the two legs, and $J'$ the interaction that couples the child spin to the two parent spins ($J'=0$ reduces the system to the ordinary 2-leg ladder); $J'$ and $J''$ are intra-household. The parameter space for strong frustration can be estimated from the $J=0$ limit, in which the system is reduced to decoupled trimers or triangles: $J''<0$ (antiferromagnetic interaction) and $|J'|\approx |J''|$. In the following, the frustration is parameterized by
\begin{equation}
\alpha=\frac{J''+|J'| }{|J|}.
\label{alpha}
\end{equation}
The smaller the magnitude of $\alpha$, the stronger the frustration. To make the ground state have much less degeneracy or entropy, the two parent spins of a household must have the same value by coupling more strongly to the child spin via $J'$ than their direct antiferromagnetic coupling $J''$, i.e., $\alpha>0$.


\subsection{The transfer matrix and SMPT}
\label{transferMatrix}
Since the model has three spins and $2^3=8$ possible states per unit cell, its transfer matrix is $8\times 8$ at a glance. However, the $m=3$ child spins can be exactly summed out as they are coupled only to the $m=1, 2$ parent spins on the same rung, yielding  \emph{the children's contribution functions}
\begin{equation}
\boxed{{}_\pm^\pm}_i=\left[\sum_{\sigma_{i,3}=\pm 1}\left(e^{\beta H^{(i)}_\mathrm{children}}\right)_{{}_{\sigma_{i,1}=\pm}^{\sigma_{i,2}=\pm}}\right]^{\frac{1}{2}}.
\label{child}
\end{equation}
They are translationally invariant, i.e., $\boxed{{}_\pm^\pm}_i=\boxed{{}_\pm^\pm}$. Then, using the spin up-down symmetry, i.e., ${\boxedmm}={\boxedpp}=[2\cosh(2\beta J')]^{1/2}$ and ${\boxedpm}={\boxedmp}=\sqrt{2}$,
we found the $4\times 4$ transfer matrix in the order of 
$\left({}_{\sigma_1}^{\sigma_2}\right) = \left({}_{+}^{+}\right), \left({}_{+}^{-}\right), \left({}_{-}^{+}\right), \left({}_{-}^{-}\right)$ to be of the same form as Eq.~(\ref{TM4simple}), with $a=e^{\beta(2J+J'')} \boxedpp^2$, $b=e^{\beta (2J-J'')} \boxedpm^2$, $u=e^{\beta (-2J+J'')} \boxedpp^2$, $v=e^{\beta (-2J-J'')} \boxedpm^2$, and $z=\boxedpp\boxedpm$. 

The largest eigenvalue of the transfer matrix is
\begin{equation}
\lambda=\Upsilon_+ \left[\cosh(2\beta J) + \sqrt{1+(\Upsilon_-/\Upsilon_+)^2\sinh^2(2\beta J)}\right],
\label{Z}
\end{equation}
with \emph{the household's frustration functions} 
\begin{eqnarray}
\Upsilon_\pm&=&e^{\beta J''}\boxedpp^2 \pm e^{-\beta J''}\boxedpm^2 \nonumber\\
&=&2e^{\beta J''}\cosh(2\beta J') \pm 2e^{-\beta J''},
\label{Y}
\end{eqnarray}
which are controlled by the intra-household interactions $J'$ and $J''$, but independent of the inter-houseld interaction $J$.
Here we have used the relationship $\Upsilon_+^2=\Upsilon_-^2+4z^2$ and $\cosh^2(x)=\sinh^2(x)+1$. 
$\lambda$ depends on $J'$ and $J''$ solely via $\Upsilon_\pm$. Note that the equations of (\ref{Z}) and (\ref{Y}) are invariant upon the transformation of $J\to -J$ or $J'\to -J'$ (i.e., ferromagnetic and antiferromagnetic interactions are interchangeable), but they are not for $J''\to -J''$. That is, only $J''<0$ (antiferromagnetic interaction) introduces frustration. 

$\Upsilon_- \propto a+u-b-v$ measures the crossing of $a+u$ and $b+v$. \textbf{It changes sign} at $T_0$ where $\Upsilon_-=0$, \red{i.e.,
\begin{eqnarray}
e^{\beta J''}\boxedpp^2 - e^{-\beta J''}\boxedpm^2 
=2e^{\beta J''}\cosh(2\beta J') - 2e^{-\beta J''} = 0.
\label{T_00}
\end{eqnarray}
This means that $T_0$ is determined only by the on-rung interactions $J'$, $J''$ and independent of the on-leg interaction $J$. For sufficiently large $2|J'|/k_\mathrm{B}T_0$, $2\cosh(2J'/k_\mathrm{B}T_0)\simeq e^{2J'/k_\mathrm{B}T_0} \gg 1$,}
\begin{eqnarray}
\label{T_0}
T_0&\simeq&\frac{2}{k_\mathrm{B}\ln2}(|J'|+J'')\\
\mathrm{for} \;\; J''<0 \;\;&\mathrm{and}&\;\; |J'|\geq|J''| \;\;(\mathrm{i.e.,}\;\; \alpha\geq 0). \nonumber
\end{eqnarray}
\red{For example, $k_\mathrm{B}T_0\approx 0.144$ for $J'=3.05$ and $J''=-3$ yielding $2\cosh(2J'/k_\mathrm{B}T_0)\approx e^{42} \gg 1$. As shown in Fig.~\ref{Fig:thermo}(a), for $J''=0.05-|J'|<0$ to keep $k_\mathrm{B} T_0\simeq 0.144$, $T_0$ as a function of $J'$ estimated by Eq.~(\ref{T_0}) (dashed lines) accurately reproduces the exact results of Eq.~(\ref{T_00}) (solid lines) except for $J'<0.15$ where $T_0$ increases exponentially; indeed, in the unfrustrated limit of $J''=0$, $T_0 \to \infty$ and no crossover would happen at finite temperature.} 

\red{It is now clear that SMPT was not found in the ordinary 2-leg ladder without the children. In that case, $J'=0$ leads to $\Upsilon_-=4\sinh(\beta J'')$, which does not change sign for given $J''$. For  $J''=0$ as the other limit of no frustration, $\Upsilon_-=4\sinh^2(\beta J')$ does not change sign, either.} 

Secondly, $(\Upsilon_-/\Upsilon_+)^2$ in Eq.~(\ref{Z}) has a prefactor of $\sinh^2(2\beta J)$, \textbf{which scales as $2^{1/\alpha}$ in the vicinity of $T_0$}. So, if Eq.~(\ref{Z}) is approximated by neglecting 1 inside $\sqrt{\cdots}$ for the strong frustration of $\alpha\ll 1$,
\begin{equation}
\lambda \simeq \Upsilon_+ \cosh(2\beta J) + |\Upsilon_-\sinh(2\beta J)|,
\label{Z2}
\end{equation}
which becomes non-analytic. The difference between Eq.~(\ref{Z}) and Eq.~(\ref{Z2}) takes place in a region of $(T_0-\delta T, T_0 +\delta T)$, where the crossover width $2\delta T$ can be estimated by $|\Upsilon_-/\Upsilon_+\sinh(2\beta J)|=1$ at $T_0\pm\delta T$, yielding
\red{
\begin{eqnarray}
2\delta T &=& \frac{2k_\mathrm{B}T_0^2}{\sinh(\frac{2J}{k_\mathrm{B}T_0})[J''+J'\tanh(\frac{2J'}{k_\mathrm{B}T_0})]} \label{dT}\\
&\simeq& \frac{8}{\ln2}T_0\;e^{-\frac{2J}{k_\mathrm{B}T_0}}\simeq\frac{8}{\ln{2}}T_0\;2^{-\frac{1}{\alpha}},
\label{dTb}
\end{eqnarray}
where Eq.~(\ref{dT}) is exact and Eq.~(\ref{dTb}) is based on Eq.~(\ref{T_0}). $2\delta T$ has two paths to approach zero: (i) $T_0 \to 0$, which is not unexpected as zero-temperature phase transition is allowed. (ii) For fixed finite $T_0$, which is determined by $J'$ and $J''$, the width $2\delta T$ approaches zero exponentially as $J$ increases. Using the above example of $k_\mathrm{B}T_0=0.144$, $2k_\mathrm{B}\delta T\simeq 1.6\times10^{-6}$ and $1.5\times 10^{-12}$ for $J=1$ and $2$, respectively. Note that $\alpha \to 0$ for both cases by Eqs.~(\ref{alpha}) and (\ref{T_0}). Although the proof of nonexistence of any phase transition in 1D Ising models with short-range interactions based on the Perron–Frobenius theorem does not work for infinite-strength interactions~\cite{Cuesta_1D_PT}, similar to zero temperature, it is amazing to realize such an ideal paradigm of SMPT in which $T_0$ and $\delta T$ are controlled by different model parameters and for fixed $T_0$, $\delta T$ decays exponentially with the single parameter $J$. This issue will be further addressed in Section~\ref{asymptotic}.}



\subsection{Thermodynamic properties}

The thermodynamical properties can be retrieved from 
the free energy per trimer $f(T)=-k_\mathrm{B}T\ln\lambda$, the entropy $S=-\partial f/\partial T$, and the specific heat $C_v=T\partial S/\partial T$.

We compare the free energies per trimer $f(T)$ obtained from using the exact Eq.~(\ref{Z}) (Fig.~\ref{Fig:thinking}b) and the mimicked Eq.~(\ref{Z2}) (Fig.~\ref{Fig:thinking}a) for $\alpha=0.05$; they differ within sub-millikelvins for $|J|=300$~K when the slope changes from near 0 to near $-k_\mathrm{B}\ln2$. The SMPT resembles a first-order phase transition with the large latent heat of $k_\mathrm{B}T_0\ln2$ per trimer.

\begin{figure}[t]
    \begin{center}
        \subfigure[][]{
\includegraphics[width=0.48\columnwidth,clip=true,angle=0]{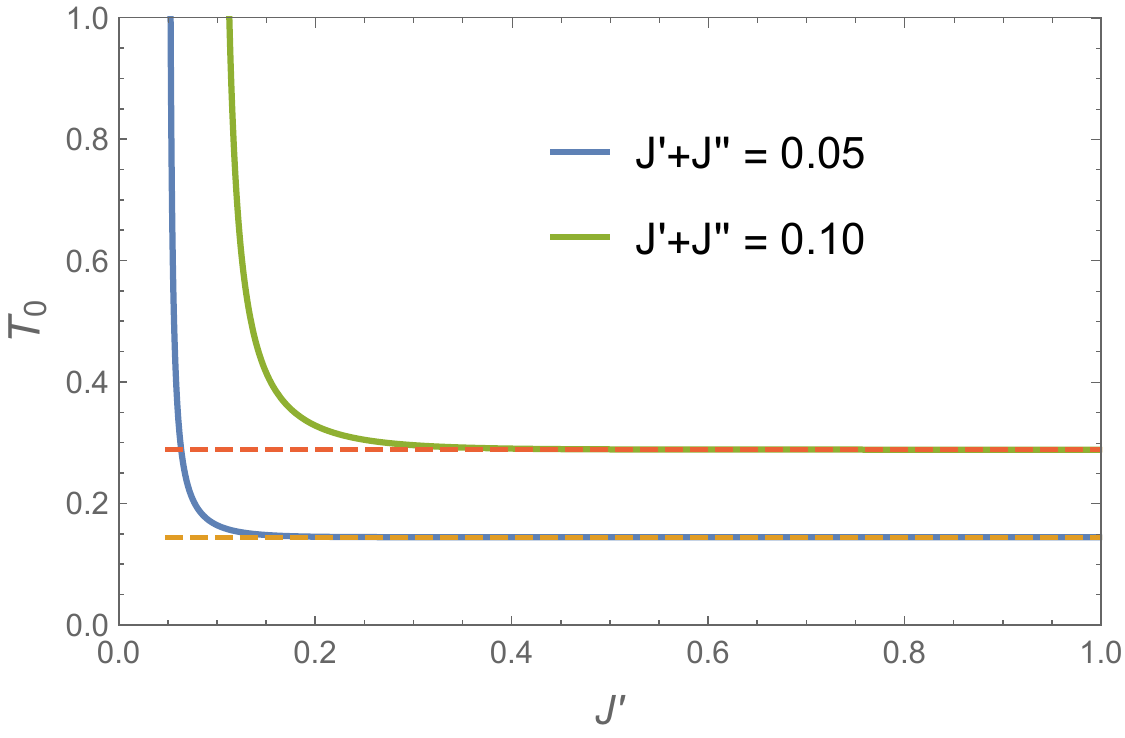}
        }
        \subfigure[][]{
\includegraphics[width=0.48\columnwidth,clip=true,angle=0]{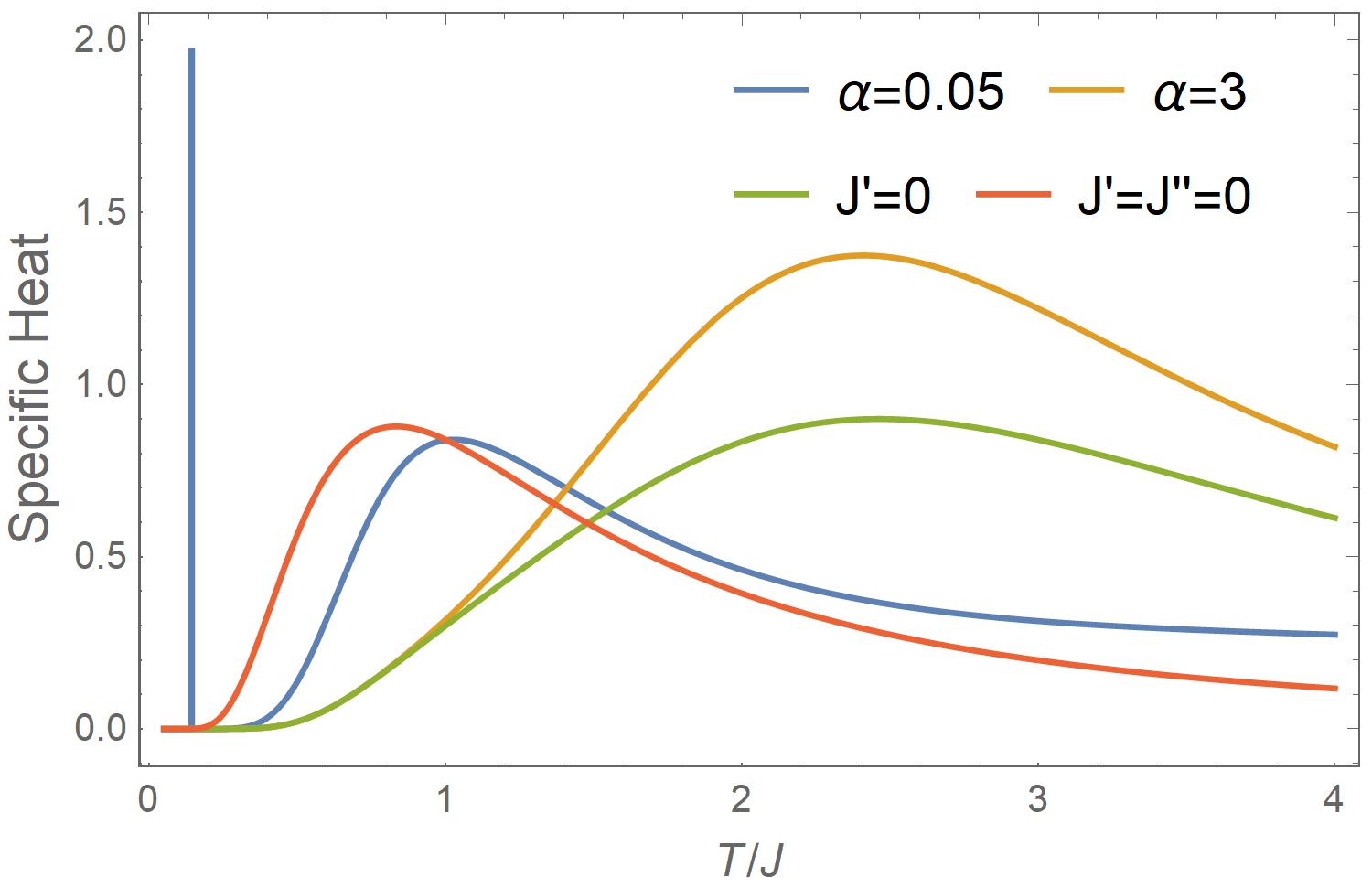}
        }
        \subfigure[][]{
\includegraphics[width=0.48\columnwidth,clip=true,angle=0]{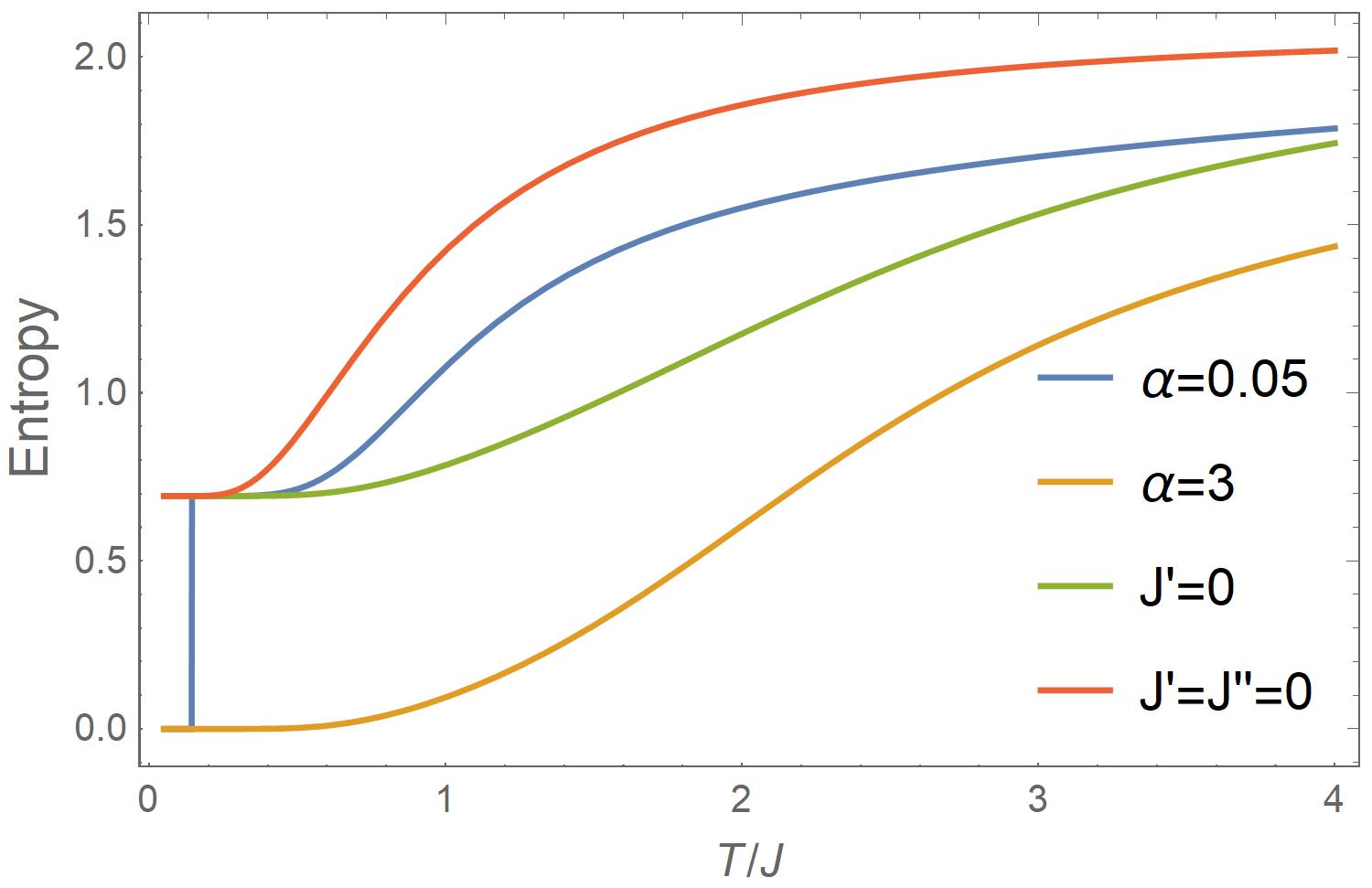}
        }
        \subfigure[][]{
\includegraphics[width=0.48\columnwidth,clip=true,angle=0]{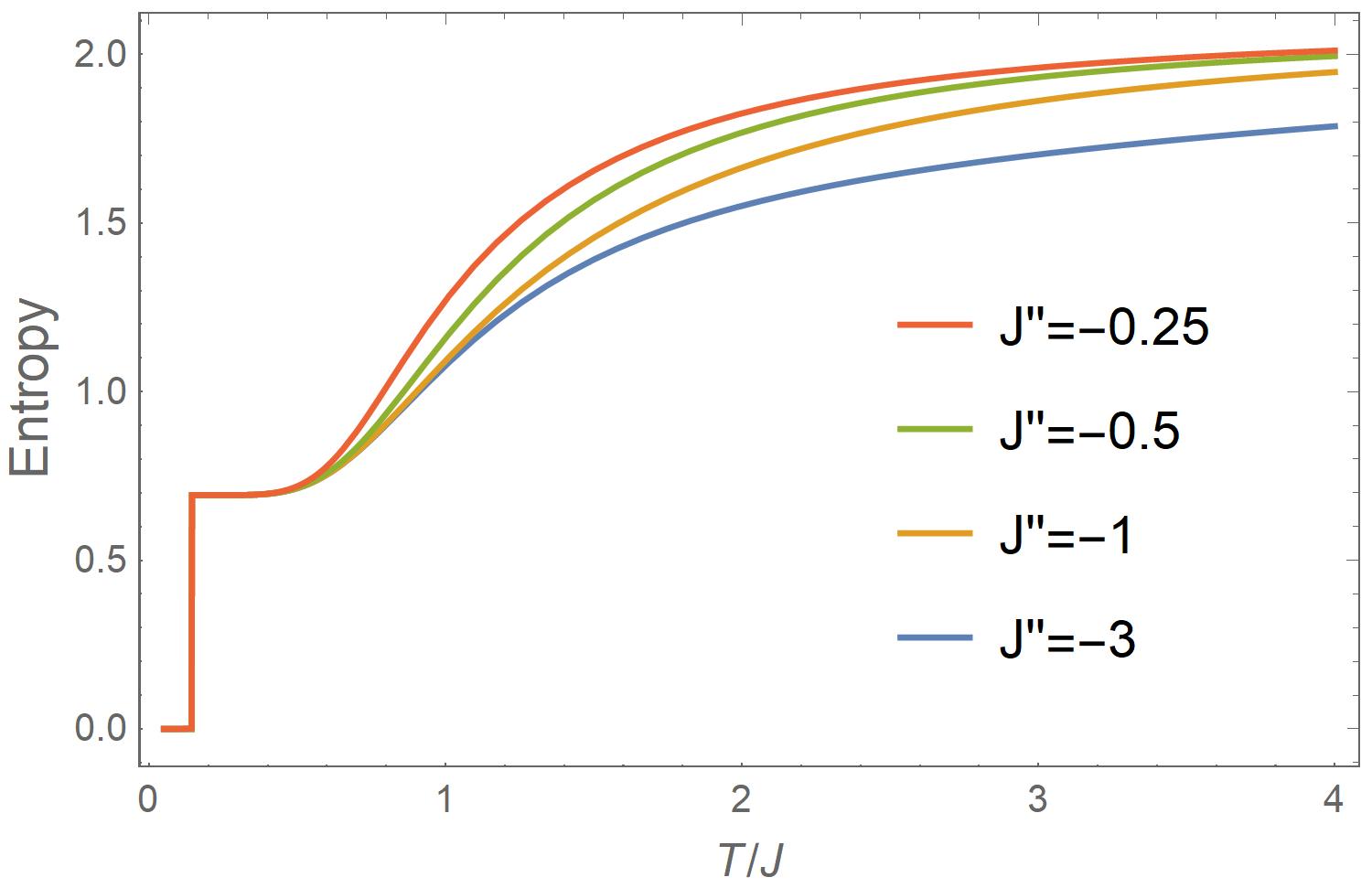}
        }
    \end{center}
\caption{\textbf{Temperature dependence of thermodynamical properties.} \red{(a) $T_0$ as a function of $J'$ with the exact results of Eq.~(\ref{T_00}) (solid lines) and the estimates by Eq.~(\ref{T_0}) (dashed lines).} (b) Specific heat and (c) entropy per trimer for four sets of model parameters. (d) Entropy per trimer for fixed $\alpha=0.05$ and four different values of $J''$. $J''=-3$ unless specified and $|J|=1$ is the energy unit.}
\label{Fig:thermo}
\end{figure}

Fig.~\ref{Fig:thermo}b shows a sharp peak in the exact specific heat for $\alpha=0.05$ (blue line), resembling a second-order phase transition. In comparison, the three typical unfrustrated cases, namely (i) $\alpha=3$ or governed by $J'$ (orange line), (ii) $J'=0$ the ordinary 2-leg ladder governed by $J''$ (green line), and (iii) $J'=J''=0$ the decoupled double chains (red line), do not show such a sharp peak.

Fig.~\ref{Fig:thermo}c show the temperature dependence of the exact entropy $S=-\partial f/\partial T$ for the same four cases as Fig.~\ref{Fig:thermo}b.
The strongly frustrated case for $\alpha=0.05$ (blue line) shows a waterfall behavior at $T_0$, where the entropy falls vertically (within $2\delta T$) from a plateau at $\ln2$ down to zero. Its low-$T$ zero-entropy behavior is in line with the unfrustrated case for $\alpha=3$ (orange line), while its intermediate-$T$ $\ln2$-plateau behavior is in line with the ordinary 2-leg ladder case for $J'=0$ (green line) where the child spins are completely decoupled from the system and thus yield the specific entropy of $\ln2$. This offers a hint to the nature of the SMPT: it is an entropy-driven transition, in which the child spins are tightly coupled to the two parent spins on the legs in the low-$T$ region because $|J'|>|J''|$, but they are decoupled from the outer ones in the intermediate-$T$ region just above $T_0$ thanks to the jumping contribution of entropy in terms of $-TS$ to the free energy. Mathematically, in the temperature region near $T_0$, one can estimate from Eq.~(\ref{Z2}) that $\lambda\approx e^{\beta (2J+2J'+J'')}$ below $T_0-\delta T$ and $\lambda\approx 2e^{\beta (2J-J'')}$ above $T_0+\delta T$, i.e., the low-$T$ region is controlled by $J'$, while the intermediate-$T$ region is controlled by $J''<0$ with the decoupled child spin contributing the factor of 2 to $\lambda$ and $\ln2$ to entropy. Moreover, in the high-$T$ region beyond the $\ln2$ plateau, the entropy does not continue following the ordinary 2-leg ladder case (green line) but behaves closer to the decoupled double chains for $J'=J''=0$ (red line). A similar behavior also shows up in the high-$T$ region of the specific heat (Fig.~\ref{Fig:thermo}a), indicating that frustration effectively decouples the chains. We shall come back to understand this strange behavior soon.

For the revealed three $T$-region behavior of the specific entropy, Eqs.~(\ref{T_0}) and (\ref{dT}) determine that the SMPT itself is a sole function of $\alpha$. This is demonstrated in Fig.~\ref{Fig:thermo}d for fixed $\alpha=0.05$ but with different $J''$, which directly couples the two legs: The lines overlap in the first two $T$-regions surrounding the waterfall at $T_0$. In contrast, in the high-$T$ region, the detail behavior of spin frustration is sensitive to $J''$ (and $J'$ shown below). It is understandable that as $|J|$ and $|J''|$ decreases, the entropy moves up to be closer to the case of the decoupled double chains. What is strange is why the chains with strong inter-chain interactions also appear to be decoupled.

\subsection{Spin-spin correlation functions and order parameters}

To find an appropriate order parameter that describes the three $T$ regions and the associated transitions, we proceed to calculate the spin-spin correlation functions, which allow us to have detailed site-by-site information:
\begin{eqnarray}
C_{mm'}(L)=\langle \sigma_{0,m}\sigma_{L,m'} \rangle_T 
= \sum_{\nu} \left(\frac{\lambda_\nu}{\lambda}\right)^L \langle \mathrm{max} | \sigma_{0,m} | \nu \rangle \langle \nu | \sigma_{L,m'}|  \mathrm{max} \rangle,
\end{eqnarray}
where $ \langle\cdots\rangle_T $ denotes the thermodynamical average, $| \mathrm{max} \rangle$ is the eigenvector of the transfer matrix corresponding to the largest eigenvalue $\lambda$ and $| \nu \rangle$ is the eigenvector corresponding to the $\nu$th eigenvalue $\lambda_\nu$. So, $C_{mm'}(\infty)=\langle \mathrm{max} | \sigma_{0,m} |  \mathrm{max} \rangle \langle \mathrm{max} | \sigma_{\infty,m'}|  \mathrm{max} \rangle$.

\begin{figure}[t]
    \begin{center}
        \subfigure[][]{
            \includegraphics[width=0.48\columnwidth,clip=true,angle=0]{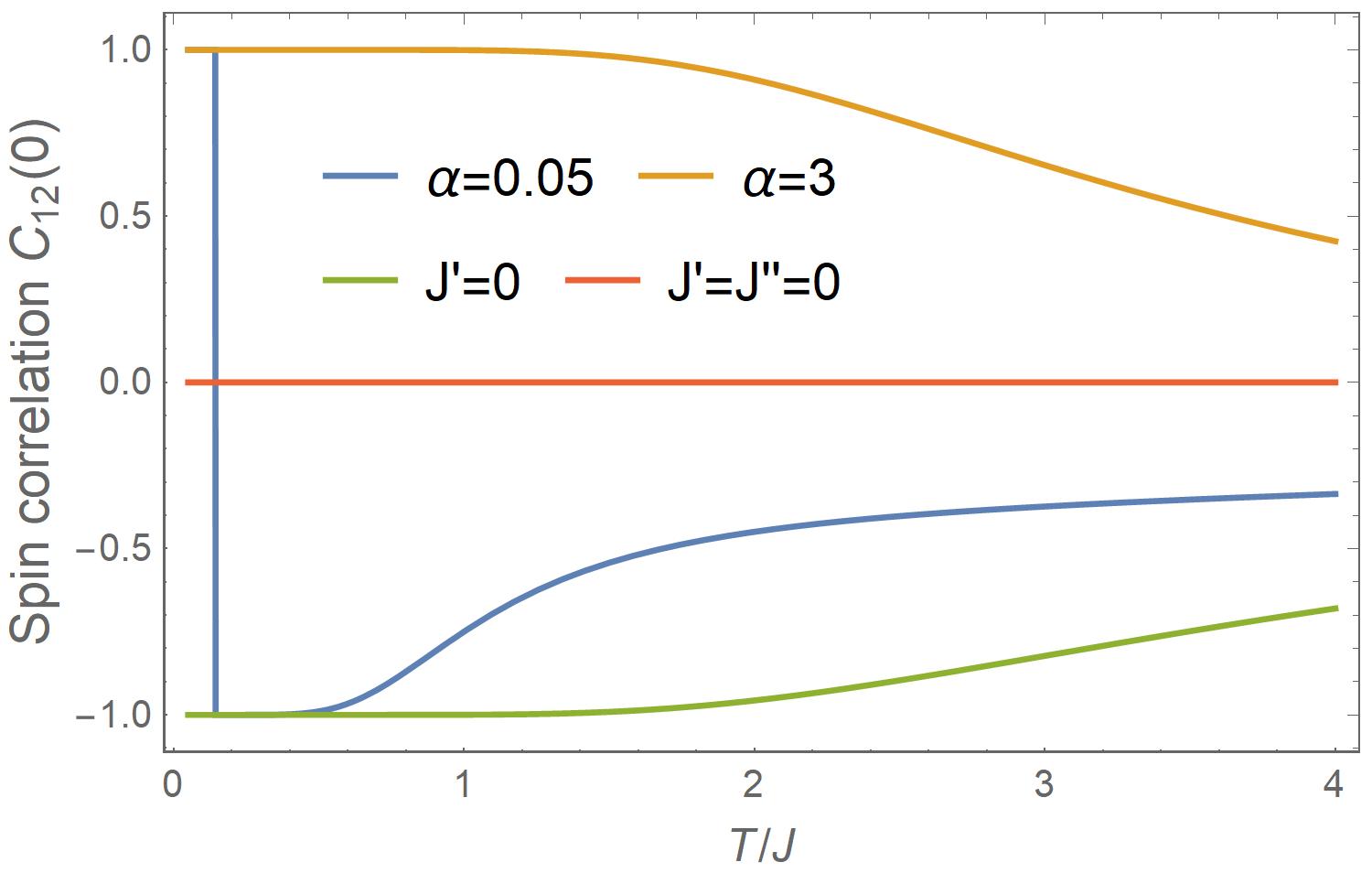}
        }
        \subfigure[][]{
            \includegraphics[width=0.48\columnwidth,clip=true,angle=0]{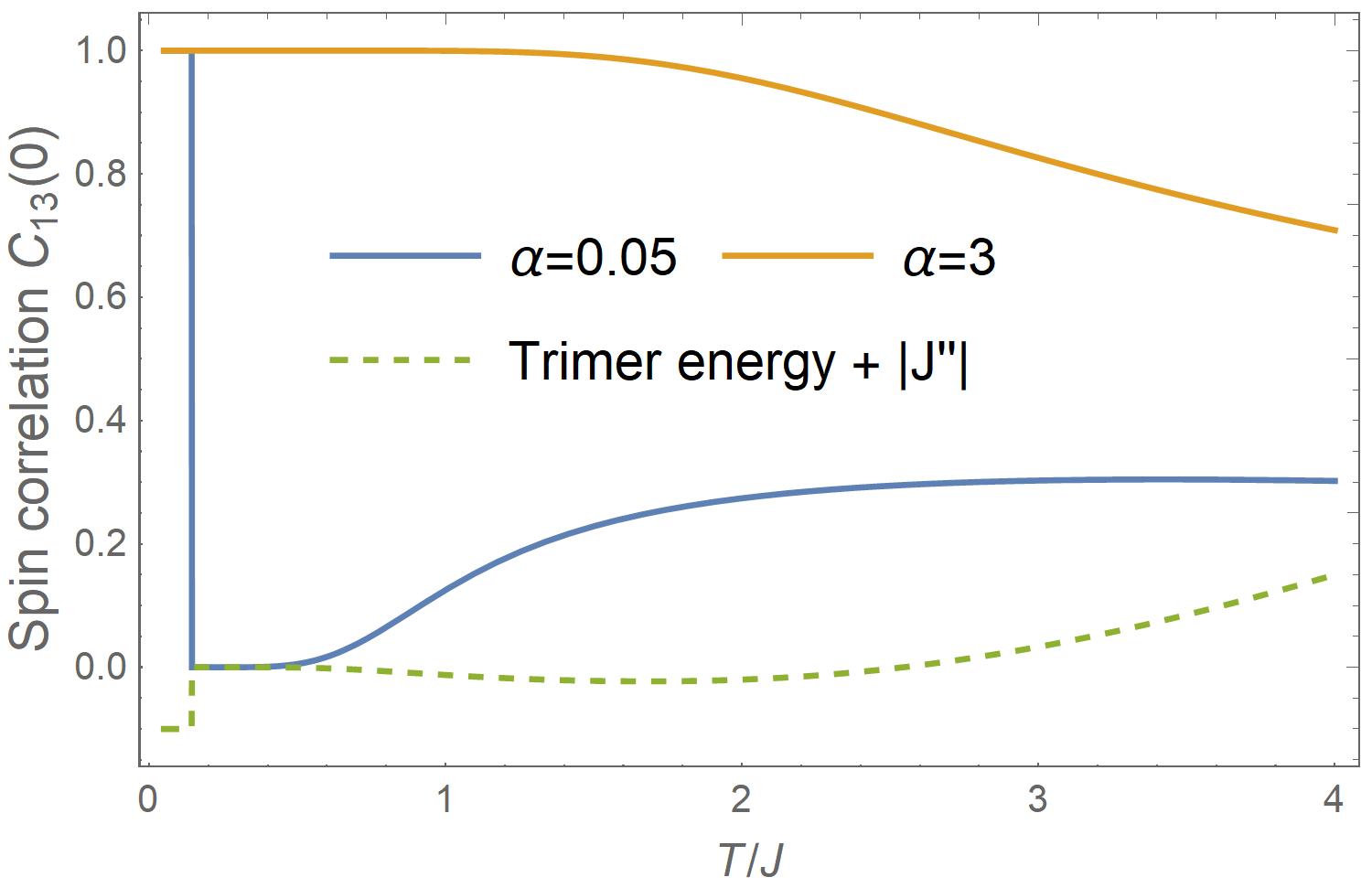}
        }
        \subfigure[][]{
            \includegraphics[width=0.48\columnwidth,clip=true,angle=0]{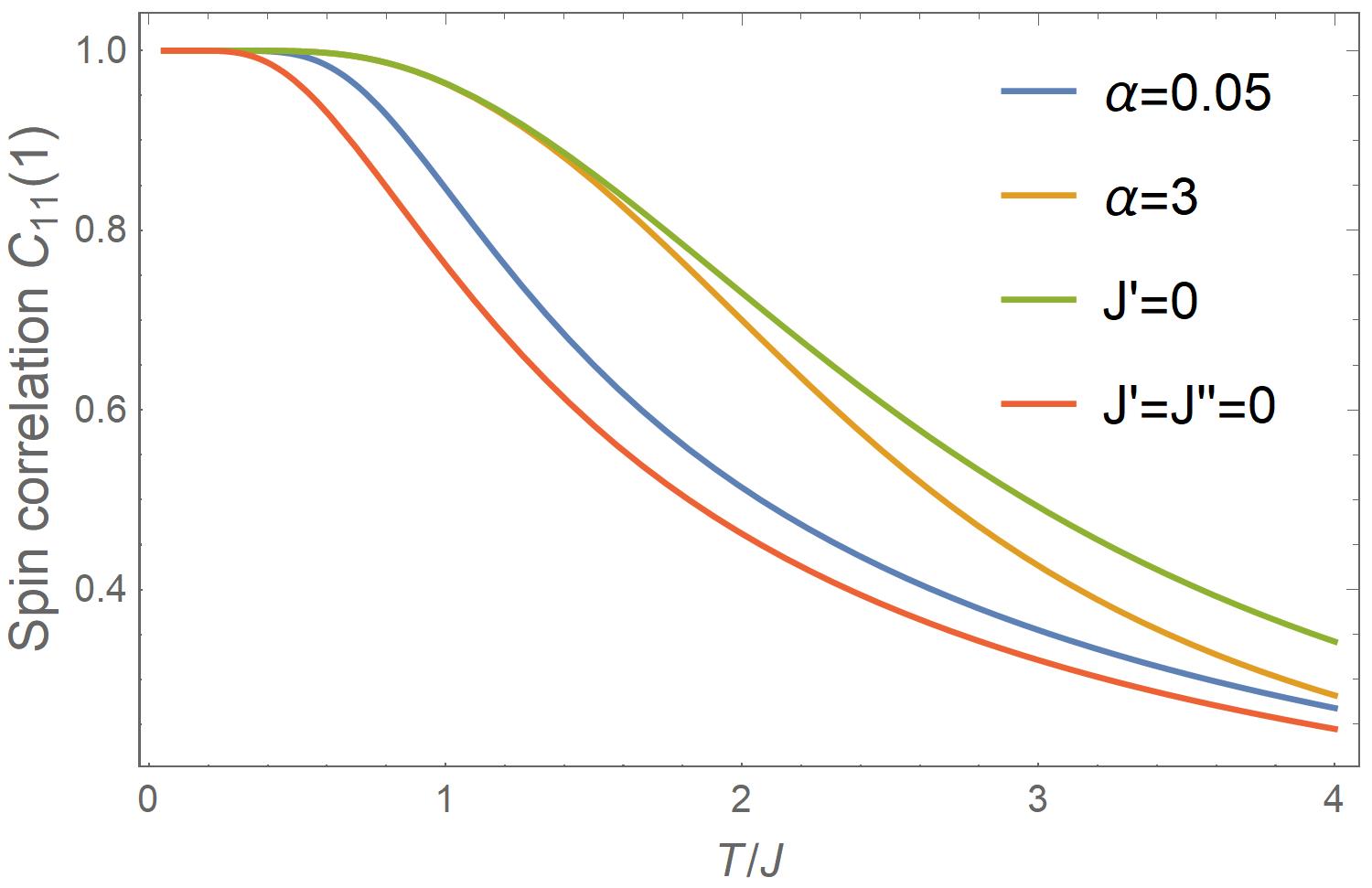}
        }
        \subfigure[][]{
            \includegraphics[width=0.48\columnwidth,clip=true,angle=0]{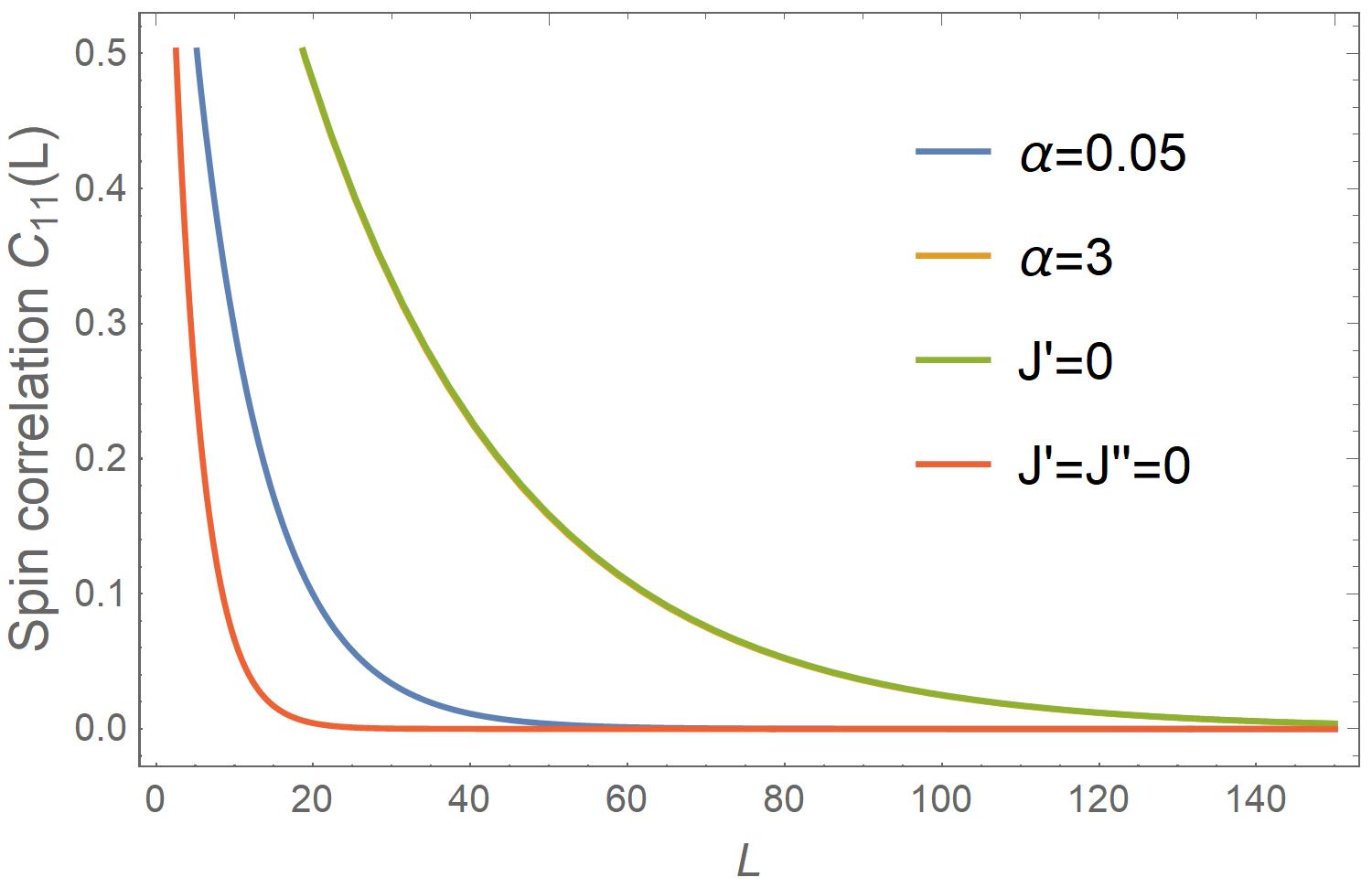}
        }
            \end{center}
\caption{\textbf{Spin-spin correlation functions and unconventional order parameters.} (a) $C_{12}(0)$, (b) $C_{13}(0)$, and (c) $C_{11}(1)$ as a function of $T$ as well as (d) $C_{11}(L)$ as a function of $L$ at $T=J$ for four sets of model parameters. The dashed line in (b) is the trimer energy $-J'' C_{12}(0)-J' C_{13}(0)-J' C_{23}(0)$, shifted up by $|J''|$ for easy reading. $J''=-3$ unless specified and $|J|=1$ is the energy unit.}
\label{Fig:SS}
\end{figure}

As shown in Figs.~\ref{Fig:SS}c and \ref{Fig:SS}d, the behavior of intra-chain correlation $C_{11}(L)$ as a function of both $T$ and $L$ (blue lines) is closer to the decoupled double chains of $J'=J''=0$ (red lines), consistent with the specific heat (Fig.~\ref{Fig:thermo}b) and the entropy data (Fig.~\ref{Fig:thermo}c). The data of the unfrustrated yet strongly coupled chains for both $\alpha=3$ (orange lines) and $J'=0$ (green lines) cases move to considerably higher temperatures, suggesting that the chain decoupling is related to the energy cost for attempts to break the order.

The conventional choice of the order parameter is the magnetization $\langle \sigma_{i,m} \rangle_T = \sqrt{C_{mm}(\infty)} = \langle \mathrm{max} | \sigma_{0,m} | \mathrm{max} \rangle = 0$ at finite $T$. This agrees with the theorems about the nonexistence of finite-$T$ phase transition in 1D Ising models with short-range interactions~\cite{Cuesta_1D_PT}. Hence, the magnetization cannot be used as the order parameter for the SMPT. Instead, let us look at the on-rung correlation function $C_{mm'}(0)= \langle \mathrm{max} | \sigma_{i,m} \sigma_{i,m'}|  \mathrm{max} \rangle$:
\begin{eqnarray}
C_{12}(0)=&\displaystyle -\frac{\partial f(T)}{\partial J''}&=\frac{(\Upsilon_-/\Upsilon_+)\cosh(2\beta J)}{\left[{1+(\Upsilon_-/\Upsilon_+)^2\sinh^2(2\beta J)}\right]^{\frac{1}{2}}}, \label{C12}\\
C_{13}(0)=C_{23}(0)=&\displaystyle -\frac{1}{2}\frac{\partial f(T)}{\partial J'}&=\tanh(2\beta J')(1+C_{12}(0))/2. \label{C13}
\end{eqnarray}
$C_{12}(0)$, the correlation between two parent spins on the legs, is proportional to the frustration function $\Upsilon_-$ and changes sign at $T_0$ (exactly zero at $T_0$). Because of the exponentially large $\cosh(2\beta J)$ and $\sinh(2\beta |J|)$ near $T_0$, $C_{12}(0)\approx \text{sgn}(\Upsilon_-) \coth(2\beta |J|) \approx +1$ below $T_0$ and $-1$ above $T_0$, as shown in Fig.~\ref{Fig:SS}a (blue line for $\alpha=0.05$). Below $T_0$, the blue line overlaps with the unfrustrated $J'$-dominated case of $\alpha=3$ (orange line). In the intermediate-$T$ region, it shows a plateau and overlaps with the unfrustrated $J''$-dominated ordinary 2-leg ladder case (green line). That is, the two parent spins changes from having like values in the low-$T$ region to having unlike values in the intermediate-$T$ region where the child spins appear to be decoupled from the ladder. So, one rung's contribution to the free energy for $T<T_0$ and $T>T_0$ is approximately $-J''-2|J'|$ and $J''-k_\mathrm{B}T\ln2$, respectively. They cross at $k_\mathrm{B}T_0=2(|J'|+J'')/\ln2=2\alpha |J|/\ln2$, in agreement with Eq.~(\ref{T_0}). What is remarkably new to learn is that for the high-$T$ region, while remaining negative, the blue line of $C_{12}(0)$ does not continue following the green line, which gradually decays to zero as temperature increases. Instead, it decays significantly faster with an inflection point at $T^*\approx 0.88$ before gradually decaying to zero. Yet, with $C_{12}(0)\approx-0.82$ at $T^*$, it is still far away from the zero red line for the case of the completely decoupled double chains. The two chains seem to be strongly coupled. How can we reconcile this with the effective decoupling seen in the above thermodynamic properties and the intra-chain correlation functions?

The answer is Eq.~(\ref{C13}) for $C_{13}(0)$ and $C_{23}(0)$, the correlations between the child spin and the two parent spins on the same rung. They are equal by symmetry. Since $C_{12}(0)$ just told us that spin 1 and spin 2 on the same rung strongly want to have unlike values, it would have been expected in static mean-field theory that $C_{13}(0)$ and $C_{23}(0)$ should have opposite signs; considering that they are required to be equal by symmetry, they should be zero. Indeed, in the intermediate-$T$ region, $C_{13}(0)$ is nearly zero (blue line in Fig.~\ref{Fig:SS}b). However, as soon as the temperature increases beyond the intermediate-$T$ region, $C_{13}(0)$ jumps up, meanwhile the magnitude of $C_{12}(0)$ decreases (blue line in Fig.~\ref{Fig:SS}a). This means that the child spin starts to re-couple to the parent spins to gain energy at the expense of $C_{12}(0)$, following the astonishing near-linear relationship of Eq.~(\ref{C13}). The total energy within one trimer $-J'' C_{12}(0)-J' C_{13}(0)-J' C_{23}(0)$ remains nearly unchanged in a wide range of temperatures (dashed line in Fig.~\ref{Fig:SS}b), which indicates that the trimers can self-organize to relieve energy cost in their value-changing dynamics. Sizable $C_{13}(0)=C_{23}(0)$ encourages the two parent spins to have like values, which counters negative $C_{12}(0)$. The net effect is decoupling the two legs.

The correlation length of the two-spin quantities $\sigma_{i,m} \sigma_{i,m'}$ is infinite at $T_0$, since the four-spin correlation function $\lim_{L\to\infty} \langle \sigma_{0,m}\sigma_{0,m'}\sigma_{L,m''}\sigma_{L,m'''} \rangle_T=C_{mm'}(0)C_{m'm''}(0)$ is not vanishing. Therefore, $C_{mm'}(0)$ are \emph{bona fide} order parameters. To check with the ordinary ladders, Figs.~\ref{Fig:SS}a and \ref{Fig:SS}b show that $C_{12}(0)$ and $C_{13}(0)$ can be finite for the other three unfrustrated cases; however, they do not change sign or they are vanishing only at $T=\infty$. This means that using $C_{mm'}(0)$ as the order parameters for the ordinary unfrustrated cases still satisfies the theorems that no phase transitions take place at finite temperature in those systems. Therefore, the unconventional order parameters $C_{mm'}(0)$ can unify the description of both the ordinary 2-leg ladder system and the unconventional phase transitions in the present frustrated ladder Ising model.

\red{To further verify that $C_{12}(0)$ is a \emph{bona fide} order parameter, we redefine the crossover width $\delta T$ by measuring the slope of $C_{12}(0)$ at $T_0$ as shown in Fig.~\ref{Fig:asymptoticity}(a): 
\begin{eqnarray}
2\delta T &=& \frac{2k_\mathrm{B}T_0^2}{\cosh(\frac{2J}{k_\mathrm{B}T_0})[J''+J'\tanh(\frac{2J'}{k_\mathrm{B}T_0})]} \label{dT2a}\\
&\simeq& \frac{8}{\ln2}T_0 e^{-\frac{2J}{k_\mathrm{B}T_0}} \simeq \frac{8}{\ln2}T_0 e^{-\frac{1}{\alpha}},
\label{dT2b}
\end{eqnarray}
which differs from the previous definition of $2\delta T$ in Eq.~(\ref{dT}) by replacing $\sinh(\frac{2J}{k_\mathrm{B}T_0})$ with $\cosh(\frac{2J}{k_\mathrm{B}T_0})$. They are consistent as $\cosh^2(\frac{2J}{k_\mathrm{B}T_0})=\sinh^2(\frac{2J}{k_\mathrm{B}T_0})+1\simeq \sinh^2(\frac{2J}{k_\mathrm{B}T_0})$ for an ultra-narrow phase crossover. Therefore, this nonclassical order parameter, which has a well-defined value space of $[-1,1]$ with the value $0$ meaning $T_0$ and its inverse slope at $T_0$ meaning $\delta T$, provides an accurate, convenient, and microscopic description of SMPT.}

The phase diagram in terms of the unconventional order parameter $C_{12}(0)$ is shown in Fig.~\ref{Fig:model}c. The frustrated 2-leg ladder Ising model can have three phases for $0<\alpha<0.15$: (i) the low-$T$ phase (red zone) is governed by $J'$ forcing the two parent spins on the same rung to have like values. (ii) The intermediate-$T$ phase (purple zone) is governed by $J''$ forcing the two parent spins on the same rung to have unlike values and the child spin to be decoupled. The abrupt switching between the two phases is an entropy-driven SMPT with large latent heat, resembling the first-order phase transition. (iii) The exotic high-$T$ phase (blue-green zone) is governed by the dynamics of the frustration. The much broader crossover between the intermediate-$T$ phase and the exotic high-$T$ phase reveals itself via the phenomenon of frustration-driven decoupling of the strongly interacted chains. The microscopic mechanism for the decoupling is illustrated in Fig.~\ref{Fig:model}c, where the high energy cost associated with the thermal activated flipping of a parent spin in the strongly interacted double chains can be relieved in the trimer dynamics by flipping the child spin cooperatively. The exact result of sizeable nonzero $C_{13}(0)=C_{23}(0)$ demonstrates that the child spin is not a slave to the mean field generated by the two parent spins, but they have equal rights. The dynamics of the trimers exhibits the art of compromise and finds the way to have created a triple-win workplace in which every neighboring bond gains an optimized share of rewards.

\red{
\section{The asymptoticity\label{asymptotic}}
}

We proceed to discuss how to arbitrarily approach the genuine phase transition at finite temperature, i.e., $\delta T \to 0$ for fixed $T_0$. This hard task becomes obvious in our model, since the value of $T_0$ is determined by $J'$ and $J''$, while the width $2\delta T$ approaches zero exponentially as $J$ increases for fixed $T_0$. Nevertheless, \emph{we hereby ask how we can improve the asymptoticity if $J$ has to be fixed or its strength cannot be further increased.} 

One simple answer is to effectively enhance $J$ by decorating the legs in such a symmetric way that the resulting $4\times 4$ transfer matrix---after the decorated spins (referred to as the \emph{bridge} sites) are summed out---is of the same form as Eq.~(\ref{TM4simple}). The convenience of using effective interactions was emphasized recently in the context of pseudo-transition~\cite{017_Krokhmalskii_PA_21_3-previous-chains_effective_model} and extended to study Eq.~(\ref{minimal}) the minimal model for SMPT~\cite{Hutak_PLA_21_trimer} soon after the model appeared in arXiv~\cite{Yin_MPT,Yin_icecreamcone}. We will show that the decoration of the legs (which controls $\delta T$) can be done independently of the decoration of the rungs (which controls $T_0$), revealing an amazing advantage of this paradigm. 

For simplicity, we consider $M$ identical bridge sites for each bond on the legs; the bridge sites do not interact with one another (Fig.~\ref{Fig:bridge}). Thus, we add the term $\Sigma_{ij} H^{(ij)}_\mathrm{bridge}$ to the model Hamiltonian Eq.~(\ref{minimal}): 
\begin{eqnarray}
H^{(ij)}_\mathrm{bridge}&=&-J_b\sum_{
k=1,2,\dots,M} 
 s_{i,j,k}(\sigma_{i,j}+\sigma_{i+1,j}),
\end{eqnarray}
where $s_{i,j,k}=\pm1$ is the Ising spin on the $k$th of the $M$ bridges connecting the $i$th and $(i+1)$th parent spins on the $j$th leg.

Like the child spins, the bridge spins can be exactly summed out, yielding \emph{the bridges' contribution functions}
\begin{equation}
\boxed{\pm\pm}_{ij}=\sum_{\substack{s_{i,j,1}=\pm 1\\ \cdots \\ s_{i,j,M}=\pm 1}}
\left(e^{\beta H^{(ij)}_\mathrm{bridge}}\right)_{{\sigma_{i,j}=\pm},\;\sigma_{i+1,j}=\pm}
\label{rainbow}
\end{equation}
They are translationally invariant and identical for both legs, i.e., $\boxed{\pm\pm}_{ij}=\boxed{\pm\pm}$. Then, using the spin up-down symmetry, i.e., ${\boxedmmh}={\boxedpph}=[2\cosh(2\beta J_b)]^M$ and ${\boxedpmh}={\boxedmph}=2^M$,
we found the $4\times 4$ transfer matrix in the order of 
$\left({}_{\sigma_1}^{\sigma_2}\right) = \left({}_{+}^{+}\right), \left({}_{+}^{-}\right), \left({}_{-}^{+}\right), \left({}_{-}^{-}\right)$ to be of the same form as Eq.~(\ref{TM4simple}), with $a=e^{\beta(2J+J'')} \boxedpp^2 {\boxedpph}^2$, $b=e^{\beta (2J-J'')} \boxedpm^2{\boxedpph}^2$, $u=e^{\beta (-2J+J'')} \boxedpp^2{\boxedpmh}^2$, $v=e^{\beta (-2J-J'')} \boxedpm^2{\boxedpmh}^2$, and $z=\boxedpp\boxedpm{\boxedpph}{\boxedpmh}$. 

We found that after introducing the following effective $J_\mathrm{eff}$ in the place of $J$,
\begin{eqnarray}
\cosh(2\beta J_\mathrm{eff}) &=& \frac{e^{2\beta J}{\boxedpph}^2 + e^{-2\beta J}{\boxedpmh}^2}{2{\boxedpph}{\boxedpmh}} \\
&=& \frac{1}{2}\left(e^{2\beta J}[\cosh(2\beta J_b)]^{M} + e^{-2\beta J}[\cosh(2\beta J_b)]^{-M}\right), \label{Jeff}
\end{eqnarray}
which is independent of the on-rung (intra-household) interactions,
the partition function remains the same except for an additional factor of ${\boxedpph}{\boxedpmh}$. This means that the household's frustration functions $\Upsilon_\pm$ are the same as defined in Eq.~(\ref{Y}) and after substituting $J_\mathrm{eff}$ for $J$, the order parameter looks the same, too:
\begin{eqnarray}
C_{12}(0)=&\displaystyle -\frac{\partial f(T)}{\partial J''}&=\frac{(\Upsilon_-/\Upsilon_+)\cosh(2\beta J_\mathrm{eff})}{\left[{1+(\Upsilon_-/\Upsilon_+)^2\sinh^2(2\beta J_\mathrm{eff})}\right]^{\frac{1}{2}}}. \label{C12eff}
\end{eqnarray}
Note that $J_\mathrm{eff} = J$ for $T\to \infty$. At the relevant low temperature, 
\begin{equation}
J_\mathrm{eff} \simeq J+M |J_b| - \frac{1}{2}M k_\mathrm{B}T\ln2,
\label{Jeff_est}
\end{equation}
that is, the effect of this bridge decoration is to enhance $J$ by about $M |J_b|$ for ferromagnetic $J>0$. As shown in Fig.~\ref{Fig:asymptoticity}c, $J_\mathrm{eff}$ as a function of $T$ estimated by Eq.~(\ref{Jeff_est}) (dashed lines) accurately reproduces the exact results of Eq.~(\ref{Jeff}) (solid lines).  

The order parameter $C_{12}(0)$ shows in Fig.~\ref{Fig:asymptoticity}b that $M=0$ with $J$ doubled (red solid line) and the decoration of $M=1$ with $J_b=J$ (green dashed line) have almost the same effect on the exponential reduction of the crossover width from $2k_\mathrm{B}\delta T\simeq 1.6\times10^{-6}$ to $1.5\times 10^{-12}$, given $k_\mathrm{B}T_0=0.144$, in agreement with Eq.~(\ref{Jeff_est}). Therefore, we can improve the asymptoticity by adding those bridges if $J$ has to be fixed or its strength cannot be further increased. To see the effect of $M$ more clearly, we present in Fig.~\ref{Fig:asymptoticity}d the results with $J_b=0.1 J$. The SMPT becomes exponentially narrower and narrower as $M$ increases, manifesting the asymptoticity of the SMPT and  making it clear that spontaneous phase transition at finite temperature does not exist in the 1D Ising model with \emph{a finite number of} short-range interactions.

The present model can be be easily extended to the case of multiple coupled children and coupled bridges. In Eq.~(\ref{child}) and Eq.~(\ref{rainbow}) for the children's and bridges' contribution functions, respectively, the detailed form of $H^{(i)}_\mathrm{children}$ or $H^{(ij)}_\mathrm{bridge}$ is unlimited. It works for arbitrary forms of interactions among the children in the same household (among the bridges between the same pair of the parents) and for both classical and quantum children/bridges---as long as their interactions with the parents are of Ising type---because the commutator $[H^{(i)}_\mathrm{children},H]=0$ and $[H^{(ij)}_\mathrm{bridge},H]=0$. So, the children/bridges can be more than spins. They can be electrons, phonons, excitons, Cooper pairs, fractons, anyons, etc. For the systems with mixed quantum particles and classical Ising spins~\cite{006_zero-field_Strecka_JMMM_16_Ising-Heisenberg-3-leg-tube,008_zero-field_Rojas_SSC_16_Ising-Heisenberg_ladder,011_Rojas_BJP_20_Ising-Heisenberg-tetrahedral_diamond,014_Rojas_JPC_20_Ising–Heisenberg_spin-1-double-tetrahedral-chain,015-7_Canova_CzechoslovakJP_04_Ising–Heisenberg_diamond_chain,015-8_Canova_JPC_06_Ising–Heisenberg-spin-S-diamond-chain}, one first obtains the eigenvalues (energy levels) of the quantum Hamiltonian $H^{(i)}_\mathrm{children}$ for one of the four $\left({}_{\sigma_1}^{\sigma_2}\right) = \left({}_+^+\right), \left({}_+^-\right), \left({}_-^+\right), \left({}_-^-\right)$ combinations of the spin configurations of the parents in the same household, say $\left({}_+^+\right)$, and thermally populates those energy levels to get ${\boxedpp}$. Then move on to work out for the other three combinations one by one. Likewise, for quantum $H^{(ij)}_\mathrm{bridge}$, work out the four combinations of the spin configurations of the nearest-neighbor parents on the same leg. Such diversity generates an infinite number of 1D systems with SMPT. The urgent questions as to how to classify nontrivial cases and reveal more and more novel effects of SMPT will be addressed in subsequent publications~\cite{Yin_icecreamcone}.

\section{Open questions}
\label{discussion}

Given the prominent roles of the Ising model and frustration in understanding collective phenomena in various physical, biological, economical, and social systems, and the prominent roles of 1D systems in research, education, and technology applications, we anticipate that the present new insights to phase transitions and the dynamical actions of frustration will stimulate further research and development about MPT. We thus leave a few open questions as our closing remarks:

\red{
(1) What are other asymptotic paths of SMPT for $h=0$? Moreover, given the present success in finding SMPT, we reconsider $h\ne 0$: what is a simple paradigm for in-field asymptotic MPT at a fixed finite temperature $T_0$? In general, the presence of the magnetic field will break the high symmetry of the $4\times 4$ matrix shown in Eq.~(\ref{TM4simple}), making it irreducible to the $2\times 2$ matrix needed for the present analysis.}  These challenges are expected to serve as a driving force for further exploration of possibilities in functionalities and their optimized performance within the huge capacity of the infinite number of 1D systems with MPT in the absence \red{or presence} of a bias (magnetic) field. 

\red{
(2) What are the first-generation SMPT-ready 1D devices for thermal applications? This appears to be feasible right now, since the Ising model has already been implemented in electronic circuits~\cite{Ising_FPGA} and optical 
networks~\cite{Pierangeli_IsingMachine_PRL19}.  
The features that $T_0$ and $2\delta T$ can be independently controlled by different parameters and different decoration methods could be attractive in engineering 1D thermal sensors, for example.  
}

(3) In the SMPT, the intermediate-temperature phase (ITP) is not a conventional, fluctuating or short-range ordered continuation of the low-temperature phase (LTP). The essence of strong frustration in driving SMPT is that the ITP is slightly higher in energy than the LTP but possesses gigantic entropy~\cite{Miyashita_10_review_frustration} as some degrees of freedom in the ITP decouple from the rest of the system. The appearance of such an ITP is reminiscent of recently discovered strange fluctuating orbital-degeneracy-lifted states at intermediate temperature in CuIr$_2$S$_4$ and other materials with active orbital degrees of freedom~\cite{Bozin_NC_CuIr2S4}. To date, 
the commonly known effect of strong frustration is to dramatically suppress $T_c$ (the critical temperature at the spontaneous phase transition) despite strong interactions, as seen in many frustrated magnets~\cite{Balents_nature_frustration}; hence, frustration has been regarded as a driver for achieving spin liquids, which do not order down to zero temperature~\cite{Balents_nature_frustration,Cao_npjQM_Ba4Ir3O10,Yin_PRL_22_trimer,Yin_PRB_21_trimer}. Could the present exact lesson encourage the use of strong frustration to understand, control, and engineer more and more LTP-ITP transitions in real materials?

(4) In a broader sense, the idea of pushing the limit in our knowledge as close as possible to the forbidden regime can be applied to other domains. For example, the Mermin–Wagner theorem~\cite{Mermin_PRL_theorem} rules out spontaneous phase transition at finite temperature in 1D or 2D isotropic Heisenberg models for quantum spins with short-range interactions. Now we ask: Does a SMPT at finite temperature exist in those quantum systems?

\begin{acknowledgments}
The author is grateful to D. C. Mattis for mailing him a copy of Ref.~\cite{Mattis_book_08_SMMS} as a gift and inspiring discussions over the years. 
Brookhaven National Laboratory was supported by U.S. Department of Energy (DOE) Office of Basic Energy Sciences (BES) Division of Materials Sciences and Engineering under contract No. DE-SC0012704.
\end{acknowledgments}

\ignore{
\section*{Appendices}
zigzag chain~\cite{Fleszar_Baskaran_J1-J2}. The one-dimensional Ising model with up to next-nearest-neighbor interactions is given by the Hamiltonian 
\begin{eqnarray}
H=-J_1\sum_{i}^{}{\sigma_{i}\sigma_{i+1}}-J_2\sum_{i}^{}{\sigma_{i}\sigma_{i+2}}.
\end{eqnarray}
The transformation $\sigma_i\sigma_{i+1} = \Tilde{\sigma}_i$, maps this model into an ordinary single-chain Ising model with magnetic field (Stephenson 1970):
\begin{eqnarray}
H=-J_1\sum_{i}^{}{\Tilde{\sigma}_i}-J_2\sum_{i}^{}{\Tilde{\sigma}_i\Tilde{\sigma}_{i+1}}.
\end{eqnarray}
}


\vspace{0.5cm}
\noindent\textbf{Author Information} The author declares no competing interests. Correspondence and requests for materials should be addressed to W.Y. (wyin@bnl.gov).




\newpage

\newpage

\newpage

\newpage

\ignore{
\newpage
\begin{figure}[t]
    \begin{center}
        \subfigure[]{
\includegraphics[width=0.55\columnwidth,clip=true,angle=0]{fig_T_alpha.png}
        }
        \subfigure[]{
\includegraphics[width=0.4\columnwidth,clip=true,angle=0]{fig_decouple.png}
        }
    \end{center}
\caption{\textbf{Phase diagram.} (a) The color map of the order parameter $C_{12}(0)$ as a function of temperature $T$ and the frustration parameter $\alpha$. Red stands for the positive $C_{12}(0)$ region governed by $J'$ (where spin 1 and spin 2 have like values), purple stands for the nearly $-1$ region governed by $J''$. They are the low-$T$ and  intermediate-$T$ phases, respectively, for strong frustration $\alpha \leq 0.15$. The blueish-to-greenish region is the exotic high-$T$ phase where frustration effectively decouples the two legs of the ladder Ising model. $J''=-3$ and $J$ is the energy unit. (b) \textbf{The mechanism of chain decoupling.} Top panel: A short-range order for $J<0$ and $J''<0$ showing alternating values on the legs and the child spins' being frustrated. Bottom panel: Spin flipping at the A site, which does not involve the child spin, costs the energy of $4|J|+2|J''|$. By contrast, spin flipping at the B site, for which the child spin responds cooperatively (illustrated for $J'<0$), costs the energy of $4|J|+2|J''|-2|J'| \approx 4|J|$, similar to the spin-flipping in the decoupled case.}
\label{Fig:phase}
\end{figure}
}

\newpage
\begin{figure}[t]
    \begin{center}
        \subfigure[]{
\includegraphics[width=0.8\columnwidth,clip=true,angle=0]{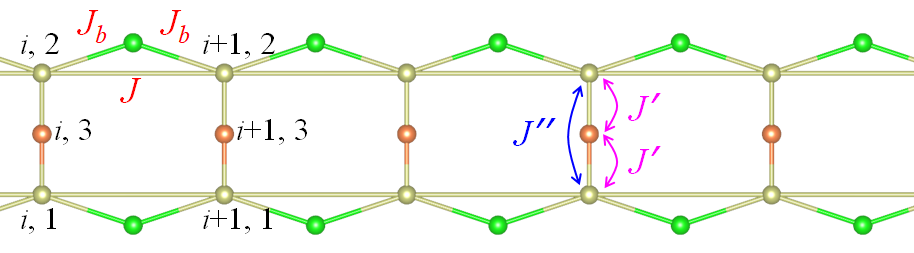}
        }
        \subfigure[]{
\includegraphics[width=0.8\columnwidth,clip=true,angle=0]{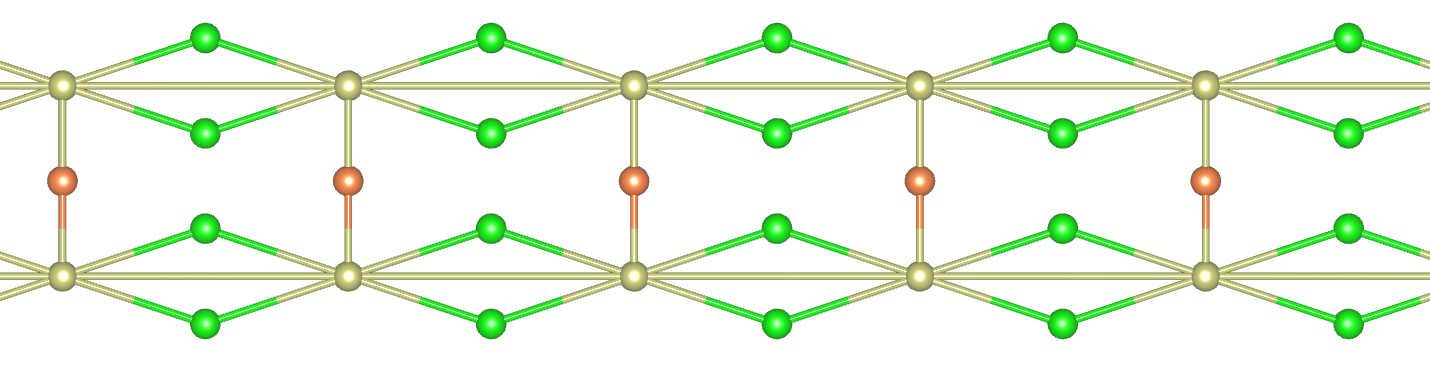}
        }
    \end{center}
        \caption{\textbf{On-leg decoration.} (a) $M=1$ and (b) $M=2$, where $M$ is the number of the decorated spins (green balls) bridging one $J$ bond connecting the nearest neighboring parent spins (yellow balls) on the legs. $J_b$ is the Ising interaction between the bridge spins and the parent spins; it effectively enhances $J$ for $J>0$ (see text).} 
\label{Fig:bridge}
\end{figure}
        
\newpage
\begin{figure}[t]
    \begin{center}
        \subfigure[][]{
\includegraphics[width=0.48\columnwidth,clip=true,angle=0]{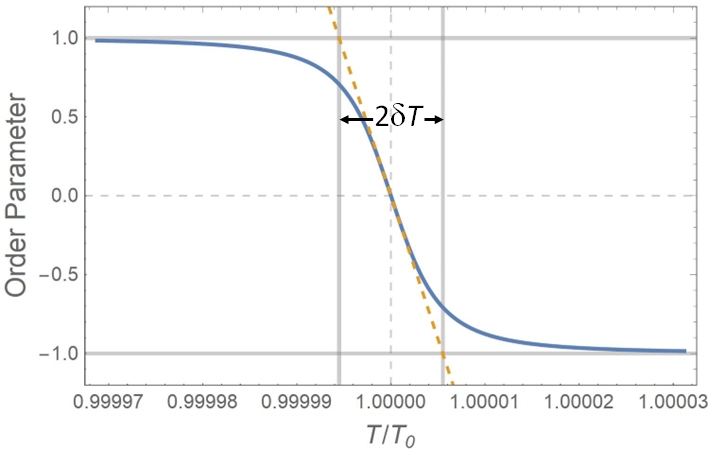}
        }
        \subfigure[][]{
\includegraphics[width=0.48\columnwidth,clip=true,angle=0]{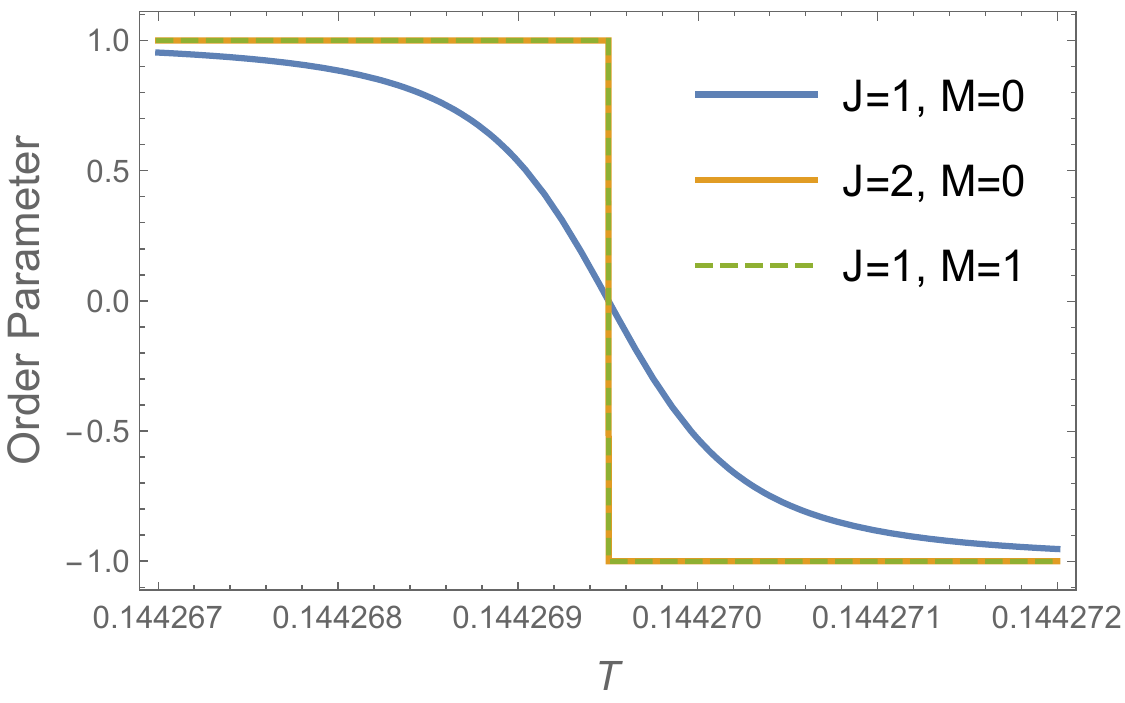}
        }
        \subfigure[][]{
\includegraphics[width=0.48\columnwidth,clip=true,angle=0]{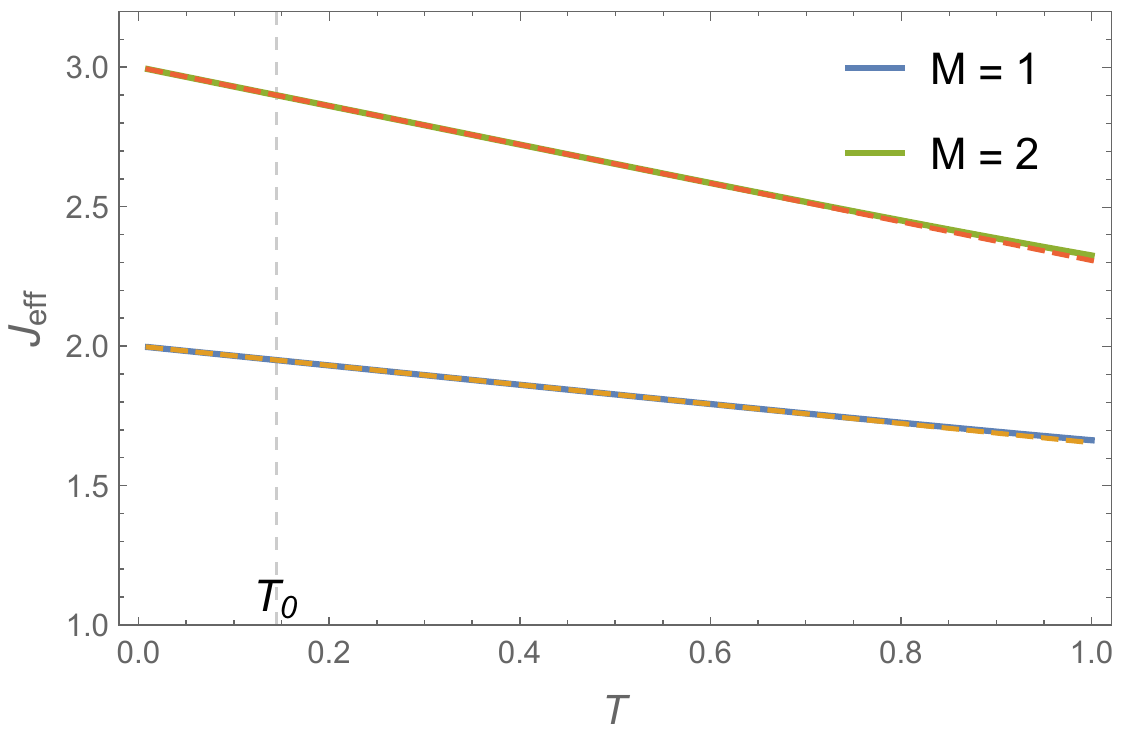}
        }
        \subfigure[][]{
\includegraphics[width=0.48\columnwidth,clip=true,angle=0]{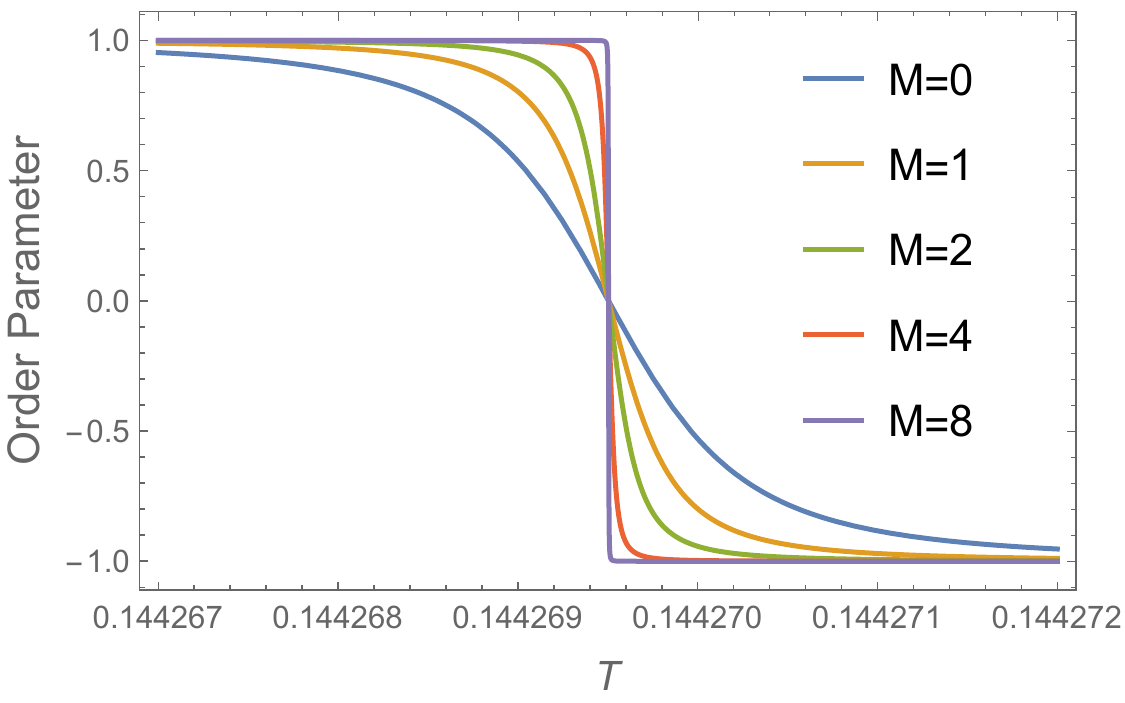}
        }
    \end{center}
\caption{\textbf{The asymptoticity of SMPT.} (a) The key measures of SMPT in terms of the order parameter $C_{12}(0)$ (blue line), where $C_{12}(0)=0$ defines $T_0$ and its inverse slope at $T_0$ (orange line) defines the width $2\delta T$.\;\; (c) $J_\mathrm{eff}$ resulting from the bridge decoration for $M=1, 2$ the number of bridges per parent bond on the legs. $J_\mathrm{eff}$ as a function of $T$ estimated by Eq.~(\ref{Jeff_est}) (dashed lines) accurately reproduces the exact results of Eq.~(\ref{Jeff}) (solid lines). Note that $J_\mathrm{eff}=J$ for $T\to \infty$.\;\;  (b)(d) The near-$T_0$ temperature dependence of $C_{12}(0)$. $k_\mathrm{B}T_0=0.144$, $J=1$ and $|J_b|=1$ except for (d) where $|J_b|=0.1$. The crossover width exponentially decreases to zero as either $J$ or $M$ increases. 
}
\label{Fig:asymptoticity}
\end{figure}

\ignore{
\newpage
\begin{figure}[t]
    \begin{center}
        \subfigure[]{
\includegraphics[width=0.45\columnwidth,clip=true,angle=0]{2D_brickwall.png}
        }
        \subfigure[]{
\includegraphics[width=0.45\columnwidth,clip=true,angle=0]{fig_model_honeycomb.png}
        }
    \end{center}
\caption{\textbf{The frustrated 2D models.} (a) Brick wall and (b) honeycomb structures with trimer vertical bonds. They are consistent with the motif that $J=0$ reduces the model to decoupled trimers. The crossover from one to two dimension can be studied with the $2n$-horizontal-chain brick-wall lattice or the $2n$-zigzag-chain honeycomb lattice with the periodic boundary condition along the $y$ direction (perpendicular to the chain direction), resulting in a nanotube-like structure. The minimal $n=1$ case is the present 2-leg ladder model with trimer rungs.}
\label{Fig:2D}
\end{figure}
}

\end{document}